\RequirePackage[utf8]{inputenc}
\pdfoutput=1
\documentclass[a4paper,aps,prx,10pt,superscriptaddress,noeprint,twocolumn, nofootinbib, longbibliography]{revtex4-1}
\usepackage{algorithm}
\usepackage{algpseudocode}
\usepackage{blindtext}
\usepackage{bm}
\usepackage{braket}
\usepackage{hyperref}
\usepackage{amssymb}
\usepackage{amsmath}
\usepackage{graphicx}
\usepackage{float}
\usepackage{bbold}
\usepackage{xcolor}
\usepackage{mathtools}
\usepackage{color,tikz,float}
\usepackage{dsfont}
\usepackage{enumitem}
\usepackage{mathrsfs}
\usepackage{stmaryrd}

\algrenewcommand\algorithmicrequire{\textbf{Input:}}
\algrenewcommand\algorithmicensure{\textbf{Output:}}

\renewcommand{\t}[1]{\textrm{#1}}
\begin{document}
\title{Quantum metrology using quantum combs and tensor network formalism}
\author{Stanis{\l}aw Kurdzia{\l}ek}
\email{s.kurdzialek@student.uw.edu.pl}

\affiliation{Faculty of Physics, University of Warsaw, Pasteura 5, 02-093 Warszawa, Poland}
\author{Piotr Dulian}
\thanks{These two authors contributed equally to the project.}
\affiliation{Faculty of Physics, University of Warsaw, Pasteura 5, 02-093 Warszawa, Poland}
\affiliation{Center for Theoretical Physics, Polish Academy of Sciences, Al. Lotnik\'ow 32/46, 02-668 Warszawa, Poland}
\author{Joanna Majsak}
\thanks{These two authors contributed equally to the project.}
\affiliation{Faculty of Physics, University of Warsaw, Pasteura 5, 02-093 Warszawa, Poland}
\affiliation{Quantum Research Center, Technology Innovation Institute, Abu Dhabi, UAE}
\author{Sagnik Chakraborty}
\affiliation{Faculty of Physics, University of Warsaw, Pasteura 5, 02-093 Warszawa, Poland}

\affiliation{Departamento de F\'isica Te\'orica, Facultad de Ciencias F\'isicas, Universidad Complutense, 28040 Madrid, Spain}
\author{Rafa{\l}  Demkowicz-Dobrza{\'n}ski} 
\affiliation{Faculty of Physics, University of Warsaw, Pasteura 5, 02-093 Warszawa, Poland}

\begin{abstract}
We develop an efficient algorithm for determining optimal adaptive quantum estimation protocols with arbitrary quantum control operations between subsequent uses of a probed channel.We introduce a tensor network representation of an estimation strategy, which drastically reduces the time and memory consumption of the algorithm, and allows us to analyze metrological protocols involving up to  $N=50$ qubit channel uses, whereas the state-of-the-art approaches are limited to $N<5$. The method is applied to study the performance of the optimal adaptive metrological protocols in presence of various noise types, including correlated noise.
\end{abstract}

\maketitle

\section{Introduction.}
One of the main lines of research in theoretical quantum metrology is development of efficient analytical and numerical tools to assess the potential of quantum probes in practical sensing scenarios \cite{Giovannetti2011, Degen2017, Pezze2018, Pirandola2018}. On the one hand this involves 
derivation of fundamental bounds on achievable sensitivity in presence of decoherence \cite{Fujiwara2008, Escher2011, Hall2012, Demkowicz-Dobrzanski2012, Demkowicz-Dobrzanski2014, Jarzyna2015, Demkowicz-Dobrzanski2017, Zhou2017, Zhou2020, Gorecki2020, Wan2022, Kurdzialek2023}, while on the other hand, development of methods that allow to directly identify the optimal metrological protocol in a particular scenario. In this paper we will focus on the latter goal.

The pursuit of identification of the optimal metrological schemes may be carried out on different levels of generality. The less fundamental, but at the same time most experimentally relevant approach, is to consider a particular physical system (light, cold atoms, etc.), consider all experimentally available degrees of freedom and resources (number of atoms, energy, time, squeezing, entanglement, ancillary systems, detectors, etc.) and come up with the scheme that yields the best sensitivity for the parameter(s) of interest.
In doing so, one may simply follow an educated-guess path, e.g. utilize squeezed states which provide reduced noise and hence better sensitivity \cite{Caves1981, Orgikh2001, Danilishin2012, Schnabel2016, Schulte2020}, or perform variational and control optimization procedures \cite{Kaubruegger2021, Yang2022, Zhang2022}. 

\begin{figure*}[t]
\includegraphics[width=.8\textwidth]{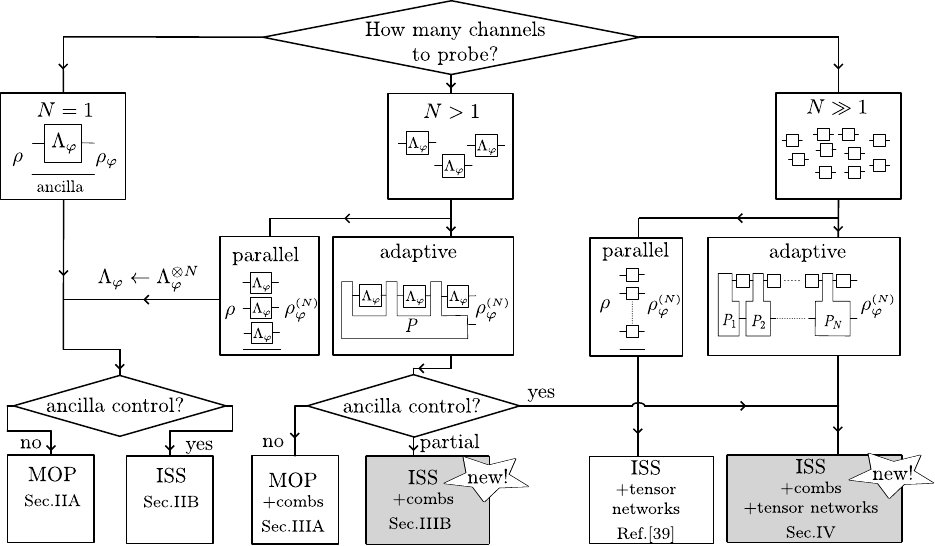}
    \caption{A big-picture view of the main results of the paper. Depending on the number of channels probed ($N$), different methods of increasing complexity should be applied. For the single channel estimation case $(N=1)$, both the minimization over purifications (MOP) method as well as the iterative see-saw (ISS) method may be directly applied. In case of relatively small $N$ (in practice $N\leq 5$ for qubit channels) both methods may be generalized to allow for the search of the optimal adaptive strategies utilizing the concept of quantum combs. For larger $N$ one is forced to employ the tensor networks techniques in order to avoid the curse of dimensionality problem, in which case the ISS method which is ideally suited. Note that in all cases, the ISS method additionally allows for an explicit control of the size of ancillary system. The bottom row of boxes serves as a guideline, indicating in which section of the paper a given method is described, with items representing the original contribution of this paper highlighted.}
\label{fig:intro}
\end{figure*}

In this paper, we take a more fundamental approach and focus on identifying metrological protocols that lead to the optimal sensing performance, irrespective of practical aspects of their implementations. This is not to say that we consider idealized, e.g. noiseless scenarios. On the contrary, we take into account imperfections and decoherence that quantum probes experience during the sensing process, but look for the optimal ways, limited only by the laws of quantum mechanics, to exploit their full quantum sensing potential---employing entanglement, quantum error-correction, active feedback, etc. 
This approach helps to understand the full sensing potential of quantum systems, and indicates space for possible improvements of existing experimental realizations. 

From this perspective, a quantum metrological problem is a quantum channel estimation problem, where the sensing process is represented by an action of a parameter(s) dependent quantum channel 
$\Lambda_\varphi$. The goal is to identify the optimal states of quantum probes as well as the measurements. This may be a reasonably easy task in case of idealized noiseless models \cite{Bollinger1996, Giovaennetti2006, Berry2000}, but becomes challenging in case of more realistic models that take into account noise and experimental imperfections \cite{Huelga1997, Dorner2009,  Knysh2014, Demkowicz-Dobrzanski2015a}. 

Fortunately, effective iterative see-saw (ISS) algorithms have been proposed that work both in the quantum Fisher information (QFI) optimization paradigm \cite{Macieszczak2013, Toth2018} as well as 
in the Bayesian one \cite{Demkowicz2011, Macieszczak2014, Jarzyna2015}. Furthermore, 
an approach of computing QFI in noisy models via minimization over purifcations (MOP) of quantum states provides an alternative method to find the optimal protocols in the form of a single semi-definite programme (SDP) \cite{Fujiwara2008, Demkowicz-Dobrzanski2012, Demkowicz-Dobrzanski2014}, see Sec.~\ref{sec:channelest} for extensive discussion.    

These approaches are effective provided the dimensionality of the quantum probe Hilbert space is small. This, in particular, makes them inefficient to use when optimizing protocols involving entangled probes that sense multiple ($N$) channels in parallel. One way this `curse of dimensionality' may be overcome is via an analysis of a particular educated-guess protocol and proving that it is the optimal one by comparing its performance with the fundamental bounds.
This was the way in which the use of squeezed light in lossy optical interferometry has been demonstrated to be asymptotically optimal \cite{Escher2011, Demkowicz-Dobrzanski2012, Demkowicz2013}, as well as the use of spin-squeezed states in Ramsey interfereomtry in presence of dephasing \cite{Orgikh2001, Ma2011, Escher2011}. 

The alternative way to face the problem in the moderate/large $N$ limit is to resort to the tensor network framework \cite{Schollwock2011}, where description of many-particle states in the form of matrix product states (MPS) has been shown to be effective in identifying the optimal states of input probes in quantum metrology \cite{Jarzyna2015}---the  method to do this is based on an appropriate reformulation of the ISS algorithm in the language of MPS and matrix product operators (MPO)  \cite{Chabuda2020, Chabuda2022}. 

From the fundamental point of view, however, the parallel sensing scheme is not the most general way to estimate a parameter encoded in a quantum channel that may be accessed a given number of times ($N$). In fact, one may consider more general adaptive protocols (e.g. quantum error-correction protocols, etc.) that admit all quantum preparation, control and measurement operations that lead to the optimal extraction of information on the parameters encoded in the quantum channels \cite{Giovaennetti2006, Demkowicz-Dobrzanski2014, Sekatski2017, Zhou2017, Zhou2020}.
The mathematical language to describe these protocols is the theory of quantum combs \cite{Chiribella2009}, within which a number of numerical methods to find optimal adaptive protocols for small scale problems have been proposed \cite{Chiribella2012, Altherr2021, Liu2023}.
These approaches were primarily based on the MOP approach  \cite{Altherr2021, Liu2023} and as such did not admit a natural way to incorporate efficient tensor network description of the protocols as well as the Bayesian approach.

This paper focuses on the development of efficient methods to identify optimal adaptive protocols in the limit of large/moderate number of channel uses $N$. In the first step we reconcile the ISS numerical approach with the quantum comb theory, see Sec.~\ref{sec:multiple_iss}. In the next step, we develop a tensor network approach that allows for an efficient identification of optimal adaptive protocols in the limit of large/moderate $N$, see Sec.~\ref{sec:tensor}. 
 The added benefits of our methods is the ability to control the effective size of ancillary systems, which is not possible in the MOP approach.  As a result, the state-of-the-art methods together with the techniques developed in this paper constitute a  comprehensive tool-box of numerical methods for quantum metrology, that may be  applied irrespective of the type of protocols analyzed and the number of channels sensed.  See Fig.\ref{fig:intro} for a big-picture view, and the context of the results presented in this paper.

Apart from these original results, we decided also to provide a comprehensive review of the MOP approaches in a unified framework  in Sec.~\ref{sec:mop} and Sec.~\ref{sec:multiple_mop} so that the paper has a self-contained character and  combines in one place many techniques that were scattered over the literature and never thoroughly discussed and contrasted with each other. These sections are not indispensable for understanding of the remaining content of the paper.

\section{Optimal channel estimation}
\label{sec:channelest}
Let us start with a paradigmatic quantum metrological problem of estimating a single parameter $\varphi$ encoded in the action of a general quantum channel $\Lambda_\varphi: \mathcal{L}(\mathcal{H}) \rightarrow \mathcal{L}(\mathcal{K})$, where $\mathcal{L}(\mathcal{H})$ represents the set of density matrices acting on Hilbert space $\mathcal{H}$, and we allow for the input and output spaces to be different. The channel is probed by an input state 
$\rho \in \mathcal{L}(\mathcal{H} \otimes \mathcal{A}$), where 
$\mathcal{A}$ represents an ancillary system with which the probing system may be entangled and on which the channel acts trivially.  The resulting output state reads $\rho_\varphi = \Lambda_\varphi \otimes \mathcal{I} (\rho)$ and is measured using a generalized measurement $\{M_i\}$. The measurement yields result $i$ with probability $p_\varphi(i) = \t{Tr}(\rho_\varphi M_i)$ and the  parameter is estimated via an estimator function $\tilde{\varphi}(i)$. The final goal is to identify the protocol where the estimated value is the closest to the true one. The exact form of the cost function depends on the approach taken, whether it is the Bayesian, min-max, or local approach involving unbiased estimators \cite{Durkin2007, Hayashi2011, Rubio2020, DemkowiczDobrzanski2020, Meyer2023}.
In this paper we will focus  on the latter approach, 
where the cost function is given by the mean squared error \begin{equation}
\label{eq:cost}
\Delta^2 \tilde{\varphi} = \sum_i p_\varphi(i) \left[\tilde{\varphi}(i) - \varphi\right]^2 \geq \frac{1}{F_Q(\rho_\varphi)},    
\end{equation}
computed at a certain operating point $\varphi$, where the estimator is assumed to be locally unbiased.
The advantage of the local approach stems from the fact that, as indicated in \eqref{eq:cost}, the cost, according to the Cram{\'e}r-Rao (CR) bound, may be lower bounded via  the inverse of the quantum Fisher information (QFI) 
$F_Q(\rho_\varphi)$, which is only a function of the output state itself \cite{Helstrom1976, Braunstein1994}. As a result, the problem of identifying the optimal estimation protocol may be reformulated
as a problem of maximization of the output state QFI:
\begin{equation}
\label{eq:maxqfi}
F_Q(\Lambda_\varphi) = \max\limits_\rho 
F_Q[\Lambda_\varphi\otimes \mathcal{I} (\rho)],
\end{equation}
where $F_Q(\Lambda_\varphi)$ is referred to as the quantum channel QFI \cite{Fujiwara2008, Kolodynski2013}. 
In what follows we will  stick to the QFI-based approach as outlined above and comment on the potential extensions to other approaches in the concluding section of the paper.  

In case of noisy channels $\Lambda_\varphi$, brute force optimization as specified in Eq.~\eqref{eq:maxqfi} via general purpose methods quickly becomes extremely inefficient even for low-dimensional systems \cite{Huelga1997, Dorner2009}. This is due to a relatively involved formula for the QFI in case of mixed states \begin{equation}
\label{eq:qfi}
F_Q(\rho_\varphi) = \t{Tr}\left(\rho_\varphi L_\varphi^2 \right), \quad \dot{\rho}_\varphi = \frac{1}{2}\{\rho_\varphi, L_\varphi\}, 
\end{equation}
where $\dot{\rho}_\varphi = \partial_\varphi \rho_\varphi$, $\{ , \}$ is the anticomutator and $L_\varphi$, implicitly defined by the right-hand-side equation above,  is the symmetric logarithmic derivative (SLD).

Over the years, two approaches proved particularly effective in solving Eq.~\eqref{eq:maxqfi} and allowed for identification of the optimal probe states in quantum metrology in regimes out of reach for general purpose optimization methods. For the sake of completeness we provide a concise review of each of these approaches below.

\subsection{Minimization over purifcations (MOP) method}
\label{sec:mop}
This method is based on the key observation that the QFI for a mixed state $\rho_\varphi$ acting on some Hilbert space $\mathcal{H}$ may be equivalently expressed as a minimization of QFI for all admissible purifcations of $\rho_\varphi$ \cite{Fujiwara2008, Escher2011, Demkowicz-Dobrzanski2012}: 
\begin{equation}
\label{eq:qfipure}
F_Q(\rho_\varphi) = \min\limits_{\ket{\Psi_\varphi}} F_Q(\ket{\Psi_\varphi}) = 4 \min\limits_{\ket{\Psi_\varphi}} \braket{\dot{\Psi}_\varphi| \dot{\Psi}_\varphi}, 
\end{equation}
where $\ket{\Psi_\varphi} \in \mathcal{H} \otimes \mathcal{R}$ is a purification of $\rho_\varphi$, $\rho_\varphi = \t{Tr}_{\mathcal{R}}\left(\ket{\Psi_\varphi}\bra{\Psi_\varphi}\right)$. 
The last equality in \eqref{eq:qfipure} is due to the fact that the explicit formula for the QFI of a pure state reads: 
$F_Q(\ket{\Psi_\varphi}) = 4 \left(\braket{\dot{\Psi}_\varphi| \dot{\Psi}_\varphi} - |\braket{\Psi_\varphi|\dot{\Psi}_\varphi}|^2 \right)$, and that given a particular purification $\ket{\tilde{\Psi}_\varphi}$ we can always find
another one $\ket{\Psi_\varphi} = e^{i \xi \varphi} \ket{\tilde{\Psi}_\varphi}$ (where $\xi=-i \braket{\dot{\tilde{\Psi}}_\varphi| \tilde{\Psi}_\varphi}$) yielding the same QFI and additionally satisfying $\braket{\Psi_\varphi|\dot{\Psi}_\varphi} = 0$. 

Utilizing this fact, we may now rewrite problem \eqref{eq:maxqfi} in the following equivalent ways \cite{Fujiwara2008}:
\begin{equation}
\label{eq:opnorm}
\begin{split}
F_Q(\Lambda_\varphi) &\overset{\t{(i)}}{=} \max_{\ket{\psi}_{\mathcal{H}\mathcal{A}}} 
    F_Q[\Lambda_\varphi \otimes \mathcal{I}(\ket{\psi}\bra{\psi})] = \\
&\overset{\t{(ii)}}{=} 4 \max_{\ket{\psi}_{\mathcal{H}\mathcal{A}}} \min_{\ket{\Psi_\varphi}_{\mathcal{H}\mathcal{A}\mathcal{R}}} \braket{\dot{\Psi}_\varphi|\dot{\Psi}_\varphi} =\\ 
&\overset{\t{(iii)}}{=} 4 \max_{\ket{\psi}_{\mathcal{H}\mathcal{A}}} \min_{\{K_{\varphi,k}\}} 
\bra{\psi} \sum_k \dot{K}_{\varphi, k}^\dagger \dot{K}_{\varphi,k}\otimes \openone_{\mathcal{A}}\ket{\psi}=\\
&\overset{\t{(iv)}}{=} 4 \max_{\rho_{\mathcal{H}}} \min_{h} \t{Tr} \left( \rho_{\mathcal{H}} \sum_k \dot{K}_{\varphi, k}^\dagger(h) \dot{K}_{\varphi,k}(h)\right) = \\
& \overset{\t{(v)}}{=} 4 \min_{h} \max_{\rho_{\mathcal{H}}} \t{Tr} \left( \rho_{\mathcal{H}} \sum_k \dot{K}_{\varphi, k}^\dagger(h) \dot{K}_{\varphi,k}(h)\right) = \\
& \overset{\t{(vi)}}{=} 4 \min_{h} \|\alpha(h) \|, \quad 
\alpha(h) = \sum_k \dot{K}_{\varphi, k}^\dagger(h) \dot{K}_{\varphi,k}(h),
\end{split}
\end{equation}
where $\| \cdot \|$ in the final formula is the operator norm.
In step (i) we make use of the fact that one may always restrict to pure input states due to convexity of the QFI. In (ii) $\ket{\Psi_\varphi}_{\mathcal{H}\mathcal{A}\mathcal{R}}$ represents a purification of the channel output state 
$\rho_\varphi = \Lambda_\varphi \otimes \mathcal{I} (\ket{\psi}\bra{\psi})$ (notice the different roles played by the ancillary system $\mathcal{A}$ and the reference system $\mathcal{R}$). In (iii) we note that an arbitrary purifcation $\ket{\Psi}$ of the state that is obtained by an action of the channel on a pure state, may be written in terms of a certain purification of the quantum channel itself, determined by a particular choice of a Kraus representation of the channel $\Lambda_\varphi$, 
$\ket{\Psi_\varphi}_{\mathcal{H}\mathcal{A}\mathcal{R}} = \sum_k K_{\varphi,k}\otimes \openone_{\mathcal{A}} \ket{\psi}_{\mathcal{H}\mathcal{A}} \otimes \ket{k}_{\mathcal{R}}$, where $\ket{k}_{\mathcal{R}}$ represents some orthonormal basis in $\mathcal{R}$.
In fact, since the quantity of interest is local (derivatives are taken at some fixed point $\varphi$), it is enough to consider a class of Kraus representations that lead to derivatives of Kraus operators of the form \cite{Escher2011, Demkowicz-Dobrzanski2012} 
\begin{equation}
\label{eq:kraush}
\dot{K}_{\varphi,k}(h) = \dot{K}^{(0)}_{\varphi,k} - i \sum_{l} h_{kl} K^{(0)}_{\varphi,l}, 
\end{equation}
where $K^{(0)}_{\varphi,l}$ is some fixed Kraus representation (e.g. canonical) and $h_{kl}$ is an arbitrary hermitian matrix. As a result, the minimization over Kraus representations effectively amounts to a minimization over a single hermitian matrix $h_{kl}$. 
In (iv) we observe that effectively the expression depends only on the reduced density matrix $\rho_{\mathcal{H}} = \t{Tr}_{\mathcal{A}}(\ket{\psi}\bra{\psi})$ and not on the whole input state $\ket{\psi}_{\mathcal{H}\mathcal{A}}$. This allows us to switch to maximization over arbitrary density matrices $\rho_{\mathcal{H}}$, which unlike pure states, form a convex set. Because of this, we can apply the minimax theorem, and switch the order of minimization and maximization (v), 
as both sets over which we optimize are convex ($h$ belongs to a linear space, $\rho_{\mathcal{H}}$ belongs to a convex set of density matrices) and the optimized function is convex in $h$ (in fact convex quadratic) and concave in $\rho_{\mathcal{H}}$ (in fact linear). Finally (vi) reflects a property of the operator norm. 

\paragraph*{Channel QFI as a semi-definite programme.}
Interestingly, the final variant for the channel QFI optimization problem can be cast as a simple semi-definite programme (SDP) \cite{Demkowicz-Dobrzanski2012}:
\begin{equation}
\label{eq:sdp}
F_Q(\Lambda_\varphi) = 4 \min_{\lambda, h} \lambda,  \ \t{subject to } A \succeq 0,
\end{equation}
where 
\begin{equation}
\label{eq:matA1}
A = \left( \begin{array}{c|ccc}
        \lambda \mathbb{1}_d& \dot{{ K}}_{\varphi,1}^\dagger(h)  & \hdots &  \dot{{ K}}_{\varphi,r}^\dagger(h) \\ \hline
        \dot{{ K}}_{\varphi,1}(h)&  &  &  	 \\
        \vdots&  & \mathbb{1}_{d \cdot r} &    \\ 
        \dot{{ K}}_{\varphi, r}(h) &  &  &            
 \end{array}\right), 
\end{equation}
where $d$ is the dimension of the space $\mathcal{H}$ and $r$ is the number of canonical Kraus operators of the channel and 
$\lambda$ is a real optimization variable. This makes this approach so appealing, as there are many good SDP solvers on the market, providing solutions accompanied by optimality benchmarks \cite{diamond2016cvxpy}, see also \cite{Boyd2004} for introduction to convex optimization and SDP in particular.

\paragraph*{Identifying the optimal input probe state.}
Interestingly, a solution of the above programme yields the desired channel QFI, but in general does not explicitly provide the form of the optimal input probe state $\ket{\psi}_{\mathcal{H}\mathcal{A}}$. Inspecting the sequence of equalities in \eqref{eq:opnorm}, one may only conclude that the reduced density matrix $\rho_{\mathcal{H}}$ corresponding to the optimal input probe state $\ket{\psi}_{\mathcal{H}\mathcal{A}}$ should be supported on the subspace spanned by eigenvectors of 
the optimal $\alpha(h)$ corresponding to the 
largest absolute eigenvalues. Only if this subspace is one-dimensional this uniquely singles out the optimal input probe state---in this case it also implies that entanglement between $\mathcal{H}$ and $\mathcal{A}$ is not required to obtain the optimal QFI.

The potential loss of information about the optimal input probe state is due to the min-max order change in step (v) in \eqref{eq:opnorm}. Let $(\rho_{\mathcal{H}}^\diamond$, $h^\diamond)$ be the optimal solution in (iv). On the other hand, let ($\rho_{\mathcal{H}}, h$) be the solution after changing the min-max order in (v). Even though, by minimax theorem $\t{Tr}[\rho_{\mathcal{H}} \alpha(h)] = \t{Tr}[\rho^\diamond_{\mathcal{H}} \alpha(h^\diamond)]$, the resulting `optimal' $\rho_{\mathcal{H}}$ will not necessarily correspond to the reduced density matrix of the optimal input state for the actual metrological task 
$\rho_{\mathcal{H}}^\diamond$ \cite{Zhou2020}. The problem is due to the fact that in the optimal solution of the min-max problem the figure of merit $\t{Tr}[\rho_{\mathcal{H}} \alpha(h)]$ can no longer in general be interpreted as the QFI of the corresponding state $\rho_{\mathcal{H}}$, as for a given $\rho_{\mathcal{H}}$ this quantity is not minimized over $h$. In order to remedy this, one needs to make sure that one is exactly at the saddle point of the function that appears under min-max, as then we are sure that for the optimal state found the corresponding $h^\diamond$ minimizes the actual figure of merit and can be regarded as the QFI of the purification of  $\rho_{\mathcal{H}}$.
Such a point always exists, and can be identified by solving the following problem: find $\rho_{\mathcal{H}} \geq 0$, $\t{Tr}(\rho_{\mathcal{H}})=1$ such that \cite{Zhou2020}:
\begin{equation}
\label{eq:saddle}
\t{Tr}\left[\rho_{\mathcal{H}} \alpha(h^\diamond)\right] = \|\alpha(h^\diamond) \|, \quad  \nabla_{h=h^\diamond} \t{Tr} \left[ \rho_{\mathcal{H}} \alpha(h)\right] =0,  
\end{equation}
where $h^\diamond$ is obtained from solving \eqref{eq:sdp}. This is again a simple SDP programme and any purification $\ket{\psi}_{\mathcal{H}\mathcal{A}}$ of such a $\rho_{\mathcal{H}}$ will correspond to the optimal input state.  

\subsection{Iterative see-saw (ISS) approach}
\label{sec:single_iss}
The MOP method is very powerful in identifying the optimal QFI and the corresponding optimal input probe state, but has at least two drawbacks, which prompt to look for alternative approaches.
The first drawback is that it is designed only for the optimization of the QFI as a figure of merit and hence is not applicable in Bayesian \cite{Macieszczak2014} or minimax \cite{Hayashi2011} (sic!) analysis. Hence, it may not be sufficient to identify optimal protocols in single-shot or finite resources regimes \cite{Gorecki2020, Rubio2020, Meyer2023, Bavaresco2023}. 
The second drawback stems from the generality of the approach, which makes it inefficient when analyzing large dimensional systems, or protocols involving multiple uses of the channels---see Section~\ref{sec:multiple}. In this approach it is in particular not possible to impose restrictions on the dimension of the ancillary system used. It is also not suitable for implementation of tensor network based protocols that allow for an efficient modelling of multiple-probe systems, as well as multiple-round adaptive strategies that avoid the `curse of dimensionality issue'---see Section~\ref{sec:tensor}. 

The alternative is the ISS procedure, that was first proposed to be used in the Bayesian phase estimation problem \cite{Demkowicz2011,Macieszczak2014}, then was generalized to deal with QFI optimization problems \cite{Macieszczak2013} and now is becoming popular in a broad range of metrological optimization tasks \cite{Toth2018, Toth2020}.   
We will present it first in the context of  QFI optimization, as this will be in fact the main focus of this paper, and only then state its Bayesian variant for completeness. For conciseness we will write $\Lambda_\varphi$ instead of more general  $\Lambda_\varphi \otimes \mathcal{I}$, which does not mean that ancilla is in general not relevant in the optimization of metrological protocols, but rather than it may always be explicitly included in the definition of the channel---this is in fact the reason why this approach, unlike the purification based method, allows to explicitly analyze the role of the ancilla and in particular its dimensionality.      

Let us start by considering the following `pre-QFI' function:
\begin{equation}
\label{eq:frl}
F(\rho,L) = 2\t{Tr}\left(\dot{\rho}_\varphi L \right)-\t{Tr}\left(\rho_\varphi L^2\right),
\end{equation}
where $\rho_\varphi = \Lambda_\varphi(\rho)$.
Maximization of the above function over Hermitian operators $L$ (since the SLD is necesserily Hermitian) yields the QFI for a given input $\rho$ \cite{Macieszczak2013}, where the corresponding optimal $L^{\diamond}$ is in fact the SLD operator as given in \eqref{eq:qfi}. This implies that we can write the channel QFI in the form of a double maximization problem:
\begin{equation}
\label{eq:doublemax}
F_Q(\Lambda_\varphi) = \max\limits_{\rho,L} F(\rho,L).
\end{equation}
This form prompts a very efficient iterative approach. We start with some random input state $\rho^{[0]}$, for which we maximize $F(\rho^{[0]},L)$ over $L$ to obtain $L^{[0]}$. Then fixing $L$ we optimize $F(\rho, L^{[0]})$ over $\rho$ to obtain $\rho^{[1]}$. We repeat this iterative procedure until $F(\rho^{[i]},L^{[i]})$ converges (e.g. does not increase by more than $0.01\%$ over five subsequent iteration steps)---the convergence to the optimal QFI value is guaranteed in generic cases , see \cite{Macieszczak2013} for the argument (to be on a safe side one should avoid choosing non-generic input states in the first step---states that are restricted to some subspace, or have certain symmetry).

The step where we find the optimal $L^{[i]}$ given $\rho^{[i]}$ amounts just to solving the linear equation for the SLD \eqref{eq:qfi}, where $\rho_\varphi = \Lambda_\varphi (\rho^{[i]})$---note that it is not advisable here to use a formula involving eigendecomposition of $\rho_\varphi$, but rather directly solve the linear equation or find $L$ as a hermitian matrix maximizing \eqref{eq:frl}, which can be formulated as an SDP. 

In order to perform the complementary step, 
note that we can rewrite the pre-QFI function as:
\begin{equation}
F(\rho,L) = 2\t{Tr}\left(\rho \dot{\Lambda}_\varphi^*(L) \right)-\t{Tr}\left(\rho \Lambda_\varphi^{*} (L^2)\right),
\end{equation}
where $\Lambda_\varphi^*(\cdot) = \sum_k K_k^\dagger \cdot K_k$ represents the dual map to $\Lambda_\varphi$. This implies that maximizing $F(\rho,L)$ over $\rho$ for a fixed $L$ amounts to solving the following problem:
\begin{equation}
\label{eq:secondstep}
\max\limits_\rho F(\rho,L) = \max\limits_\rho \t{Tr}\left(\rho M \right), \ M = 2 \dot{\Lambda}_\varphi^*(L) -  \Lambda_\varphi^{*} (L^2),
\end{equation}
with standard constraints on the input state $\rho \geq 0$, $\t{Tr}( \rho) = 1$. This is a simple SDP programme, for which the solution can be written explicitly as $\rho= \ket{\psi^+}\!\bra{\psi^+}$, where $\ket{\psi^+}$ is the eigenvector of $M$ corresponding to its largest  
eigenvalue.

There are a number of variations of the algorithm, where one may restrict the set of allowed input states $\rho$ to some convex set, or fix the measurement, in which case the optimized quantity is the classical Fisher information (FI) \cite{Macieszczak2013}. Most importantly, the procedure may be easily adapted to minimize the Bayesian quadratic cost with arbitrary prior distribution for the estimated parameter $p(\varphi)$, which we briefly review below.  

The average Bayesian quadratic cost for estimating parameter $\varphi$ is defined as
\begin{equation}
\label{eq:bayesiancost}
\overline{\Delta^2\tilde{\varphi}} =  \int \t{d}\varphi\, p(\varphi) \sum_i  p_\varphi(i) \left[\tilde{\varphi}(i) - \varphi\right]^2
\end{equation}
and corresponds to \eqref{eq:cost} averaged over the prior. Minimization of the cost over measurements, estimators and input states results in the formula for the minimal Bayesian cost of estimating a parameter of the channel in the form \cite{Macieszczak2014, Jarzyna2015}:
\begin{equation}
\overline{\Delta^2\tilde{\varphi}}(\Lambda_\varphi) = \Delta^2 \varphi - \max \limits_{\rho,\bar{L}} \t{Tr}\left( 2 \bar{\rho}^{\prime} \bar{L}  - \bar{\rho} \bar{L}^2  \right),
\end{equation}
where $\Delta^2 \varphi$ represents variance of the prior distribution $p(\varphi)$, 
$\bar{\rho}=   \int \t{d} \varphi \, p(\varphi) \rho_\varphi$ is the output state averaged with respect to the prior, whereas 
$\bar{\rho}^\prime = \int \t{d} \varphi \, p(\varphi) (\varphi - \bar{\varphi})\rho_\varphi$ with $\bar{\varphi}$ being the prior expectation value of $\varphi$. Comparing the above formula with (\ref{eq:frl},\ref{eq:doublemax}) we see that an analogous iterative optimization procedure may now be applied. With fixed $\rho$, the search for the corresponding optimal $\bar{L}$ amounts to solving the SLD-like equation, \eqref{eq:qfi}, with $\rho_\varphi$ replaced by $\bar{\rho}$ and $\dot{\rho}_\varphi$ replaced by $\bar{\rho}^\prime$.       On the other hand, with fixed $\bar{L}$ the search for the optimal input state $\rho$ amounts to the search for the eigenstate corresponding to the largest eigenvalue of
\begin{equation}
\bar{M} = \int \t{d} \varphi\, p(\varphi)
\Lambda_\varphi^*\left[ 2 (\varphi - \bar{\varphi}) \bar{L}  - \bar{L}^2 \right].
\end{equation}
This shows how versatile the see-saw method is, as it can equally well address two conceptually different estimation problems.

\section{Optimal channel estimation with multiple coherent uses}
\label{sec:multiple}
In the previous section we have presented efficient methods to identify optimal metrological protocols focusing on a single use of a quantum channel. In quantum metrology, however, we most typically face a situation where quantum channels may be utilized multiple times. Physically, this represents situations when we are able to utilize many quantum probing systems simultaneously (e.g. multiple atoms sensing common magnetic field, multiple photons travelling through the same interferometer, etc.), or utilize a single quantum system 
to perform a coherent sensing of the same environment over extended periods of time  (single photon bouncing multiple-times in a cavity, single spin experiencing the same magnetic field over long times, etc.) or a combination of both. In order to understand the fundamental potential of quantum metrology, one should be able to identify the optimal protocols which lead to the best estimation of quantum channel parameters, for a given number of channel uses, assuming any kind of quantum control is allowed.

One option is to probe quantum channels in parallel, sending (possibly) entangled states of $N$ probes into $\Lambda_\varphi^{\otimes N}$. Note that this situation is formally equivalent to the one discussed in the previous section, where $\Lambda_\varphi$ needs to be replaced by $\Lambda_\varphi^{\otimes N}$. Hence, the methods presented above are valid, and can be applied provided $N$ is not too large, as in this case the curse of dimensionality makes the problem numerically intractable. If this is actually the case, however, one may no longer optimize over arbitrary input states but instead needs to restrict to some reasonable classes of states and compute their performance without invoking the formalism of the full Hilbert space.  This can be done using tensor networks methods \cite{Chabuda2020, Chabuda2022}, or by direct Heisenberg picture computations of the performance of particular input probe states and measurement observables \cite{Demkowicz-Dobrzanski2015a, Ma2011}. When the performance of the protocols is shown to coincide with some fundamental bounds, this proves the protocols are indeed optimal.  

Still, the parallel schemes do not cover the most general quantum adaptive strategies, including quantum control, active quantum feedback, quantum error correction, etc. 
\begin{figure}[t]
\includegraphics[width=1.\columnwidth]{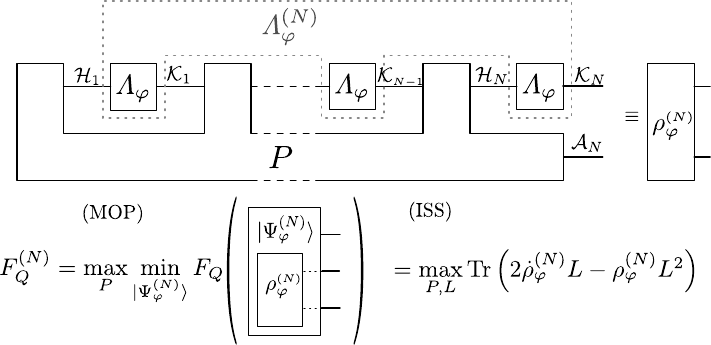}
    \caption{
        A general scheme of coherent probing of
        $N$ independent quantum channels $\Lambda_\varphi$
        (or a general global channel $\Lambda^{(N)}_\varphi$---gray) via an adaptive strategy represented by a quantum comb $P$. Quantum Fisher Information of the output state is optimized over $P$ either via minimization over purification (MOP) method or the iterative see-saw approach (ISS).}
\label{fig:comb}
\end{figure}
The problem of the search for the most general quantum adaptive strategy may be formulated as follows. 
Given $N$ uses of a quantum channel $\Lambda_\varphi: \mathcal{L}(\mathcal{H}) \rightarrow \mathcal{L}(\mathcal{K})$ find the 
optimal superoperator $\textrm{P}$, see Fig.~\ref{fig:comb}, that 
yields the output state $\rho^{(N)}_{\varphi} \in \mathcal{L}(\mathcal{K}_N \otimes \mathcal{A}_N)$ with maximal possible QFI (or minimizing the corresponding Bayesian cost if one follows the Bayesian approach). 
In what follows, we will denote superoperators with non-italic font $\textrm{P}$, while its italic variant $P$ will represent the corresponding  Choi-Jamio{\l}kowski (CJ) operator \cite{Bengtsson2006}, i.e.:
\begin{equation}
P = \textrm{P} \otimes \mathcal{I}_{\mathcal{H}_{\t{in}}} \left(\ket{\Phi}\bra{\Phi} \right),	
\end{equation}
where $\mathcal{I}_{\mathcal{H}_{\t{in}}}$ is the identity operator on tensor product of all input spaces $P$ (in our case $\mathcal{H}_{\t{in}} = \mathcal{K}_1 \otimes \dots \mathcal{K}_{N-1}$) while $\ket{\Phi}= \sum_i \ket{i}\otimes \ket{i}$ is a non-normalized maximally entangled state on $\mathcal{H}_{\t{in}} \otimes \mathcal{H}_{\t{in}}$, where $\ket{i}$ is an orthonormal basis in $\mathcal{H}_{\t{in}}$.  
In particular, 
$\varLambda_\varphi \in \mathcal{L}(\mathcal{K}\otimes \mathcal{H})$ will represent the CJ operator of $\Lambda_\varphi$.

The superoperator $\t{P}$ represents all possible interaction of the sensing system with arbitrarily large ancillary systems, and allows for any control operations in between subsequent channel uses. Mathematically, $\t{P}$ is a linear operator $\t{P}: \mathcal{L}(\mathcal{K}_1 \otimes \dots \mathcal{K}_{N-1}) \rightarrow 
\mathcal{L}(\mathcal{H}_1 \otimes \dots \mathcal{H}_{N}\otimes \mathcal{A}_N)$ 
that satisfies conditions for being a quantum comb \cite{Chiribella2009} 
$\t{P} \in \t{Comb}[(\emptyset, \mathcal{H}_1), (\mathcal{K}_1, \mathcal{H}_2),...,(\mathcal{K}_{N-1}, \mathcal{H}_{N} \otimes \mathcal{A}_N)]$, where pairs of spaces represent respective input/output spaces of each `tooth' of the comb. 
In terms of the corresponding CJ matrix $P \in \mathcal{L}(\mathcal{H}_1\otimes \mathcal{K}_1 \otimes \dots \mathcal{H}_{N-1} \otimes \mathcal{K}_{N-1} \otimes \mathcal{H_N} \otimes \mathcal{A}_N)$ (the ordering of spaces is chosen for notational convenience) the conditions for $\t{P}$ being a comb read:
\begin{align}
\label{eq:combcons}
&P\geq 0, \ \t{Tr}_{\mathcal{A}_N \otimes \mathcal{H}_N} {P} = {P}^{(N-1)} \otimes \openone_{\mathcal{K}_{N-1}}, \\ 
\nonumber
&\underset{1 <  k<N}{\forall} \t{Tr}_{\mathcal{H}_k} {P}^{(k)} = {P}^{(k-1)} \otimes \openone_{\mathcal{K}_{k-1}}, \ \t{Tr}_{\mathcal{H}_1} {P}^{(1)}=1, 
\end{align}
where $P^{(k)}$ represents ${P}$ traced out over $\mathcal{K}_k \otimes \mathcal{H}_{k+1}\otimes \dots\otimes \mathcal{K}_{N-1}\otimes \mathcal{H}_{N}\otimes \mathcal{A}_N$. Intuitively, these conditions are related to the causal structure of adaptive protocols---input $\mathcal{K}_i$ may affect only  outputs $\mathcal{H}_{i+1},...,\mathcal{H}_N$, all previous outputs cannot depend on $\mathcal{K}_i$, see Ref.~\cite{Chiribella2009} for a further discussion.
The final output state $\rho_\varphi^{(N)}$ is now obtained by concatenating $\t{P}$ operation with 
$N$-fold use of the channel $\Lambda_\varphi$, which mathematically corresponds to application of the link product operation to the corresponding CJ operators \cite{Chiribella2009}, defined as follows: 
\begin{equation}
\label{eq:link}
\rho_{\varphi}^{(N)} = {\varLambda}_\varphi^{\otimes N} \star {P}.
\end{equation}
Given two operators ${A} \in \mathcal{L}(\mathcal{A}\otimes \mathcal{C})$, ${B} \in \mathcal{L}({ \mathcal{C} \otimes\mathcal{B} })$, where the common subsystem on which they act is denoted by $\mathcal{C}$, the link product is defined as ${A} \star {B} = \t{Tr}_\mathcal{C}\left[({A} \otimes \openone_{\mathcal{B}}) (\openone_{\mathcal{A}} \otimes {B}^{T_{\mathcal{C}}})  \right]$, where $T_\mathcal{C}$ denotes transposition with respect to subsystem $\mathcal{C}$. Note that when ${A}=\ket{A}\!\bra{A}$, ${B} = \ket{B}\!\bra{B}$ are rank-1 operators, so is their link product, as ${A} \star {B} = \ket{A \star B}\bra{A \star B}$, where $\ket{A \star B} := \sum_{c} \braket{c|A} \otimes \braket{c|B}$, with $\{\ket{c} \in \mathcal{C}\}$ representing the basis in $\mathcal{C}$ (the distinguished basis, with respect to which the partial transposition is defined).

The problem of identifying the optimal metrological protocol now amounts to the following optimization task:
\begin{equation}
\label{eq:maxqficomb}
F^{(N)}_Q(\Lambda_\varphi) = \max\limits_{{P}} 
F_Q[{\varLambda}_\varphi^{\otimes N} \star {P}],
\end{equation}
with constraints on ${P}$ given in \eqref{eq:combcons}.
When compared with the single channel estimation problem, \eqref{eq:maxqfi}, the only difference amounts to the replacement of the input state $\rho$ with the quantum comb $P$. This problem can again be approached using either the MOP or the ISS method. In the discussion below, we will consider an even more general scenario, where we replace $N$ independent uses of a channel $\Lambda_\varphi$ by an arbitrary $N$-teeth quantum comb $\Lambda_\varphi^{(N)}$---mathematically this amounts to replacing $\Lambda_\varphi^{\otimes N}$ by $\Lambda_\varphi^{(N)}$. 
This will allow us to also discuss the models where the probed channels are subject to e.g. correlated noise, see Fig.~\ref{fig:comb} and Section \ref{sec:corr_deph}.

\subsection{Minimization over purifcations (MOP) method}
\label{sec:multiple_mop}
Since the set of quantum combs is convex and the output state $\rho_{\varphi}^{(N)}$ is a linear function of ${P}$, the  convexity of QFI implies that we may restrict ourselves to extremal quantum combs at the input. 
Unlike in the single-channel estimation case, an extremal comb is not necessarily pure, i.e. a CJ operator of rank one $P = \ket{P}\!\bra{P}$, due to non-trivial constraints \eqref{eq:combcons}. Still, we may always purify it  at the expense of possibly increasing the dimension of the ancillary system $\mathcal{A}_N$. Hence, we may now follow an analogous procedure as in \eqref{eq:opnorm}, by noticing that 
the minimization of the QFI formula over different purifications of $\rho_{\varphi}^{(N)}$ can again be understood in terms of minimization over different Kraus representations of $\Lambda_{\varphi}^{(N)}$.  
Let $\{K_{\varphi,k}^{(N)}\}$ be a Kraus representation of $\Lambda_{\varphi}^{(N)}$, which can be written in terms of the decomposition of the corresponding CJ matrix  ${\varLambda}_{\varphi}^{(N)} = \sum_k \ket{K_{\varphi,k}^{(N)}}\!\bra{K_{\varphi,k}^{(N)}}$, where $\ket{K_{\varphi,k}^{(N)}} \in \mathcal{K}_1 \otimes \mathcal{H}_1 \otimes \dots \otimes \mathcal{K}_N \otimes \mathcal{H}_N$ represents a vectorized Kraus operator. Each Kraus representation may be associated with the following purification  
   $\ket{\Psi^{(N)}_\varphi}_{\mathcal{K}_N \mathcal{A}_N \mathcal{R}} = \sum_k \ket{ K_{\varphi,k}^{(N)} \star P} \otimes \ket{k}_{\mathcal{R}} \in \mathcal{K}_{N}\otimes \mathcal{A}_N \otimes \mathcal{R}$, where
   we have used the notation for the link product of rank-1 operators. 
   With this notation, we can now adapt \eqref{eq:opnorm} in order to derive the formula for the optimal QFI: 
\begin{equation}
\label{eq:compopt}
\begin{split}
F_Q(\Lambda^{(N)}_\varphi) &=  
 4 \max_{{P}} \min_{\ket{\Psi^{(N)}_\varphi}_{\mathcal{K}_N\mathcal{A}_N \mathcal{R}}} \braket{\dot{\Psi}^{(N)}_\varphi|\dot{\Psi}^{(N)}_\varphi} =\\ 
&=  4 \max_{{P}} \min_{\{K^{(N)}_{\varphi,k}\}} 
\sum_k  \braket{\dot{K}_{\varphi,k}^{(N)} \star P| \dot{K}_{\varphi,k}^{(N)} \star P} = \\
&=  4 \max_{{P}} \min_{\{K^{(N)}_{\varphi,k}\}} \t{Tr} \left( \sum_k \ket{\dot{K}^{(N)}_{\varphi,k}}\!\bra{\dot{K}^{(N)}_{\varphi,k}}\star {P}\right) = \\
& =   4 \max_{{P}} \min_{h} \t{Tr} \left[( \Omega(h) \otimes \openone_{\mathcal{A}_N}) ({P} \otimes \openone_{\mathcal{K}_N})\right ] = \\
& \overset{\t{(i)}}{=}  4 \max_{{\tilde{P}}} \min_{h} \t{Tr} \left[\tilde{\Omega}(h) {\tilde{P}}\right ] =\\
& \overset{\t{(ii)}}{=} 4 \min_{h} \max_{{\tilde{P}}} \t{Tr} \left[\tilde{\Omega}(h)  {\tilde{P}}\right ],
\end{split}
\end{equation}
where $\Omega(h) = \sum_k\ket{\dot{K}^{(N)}_{\varphi,k}(h)}\!\bra{\dot{K}^{(N)}_{\varphi,k}(h)}^T$, while $\ket{\dot{K}^{(N)}_{\varphi,k}(h)}$ is defined analogously as in \eqref{eq:kraush}, but this time $h$ is in principle a huge matrix, as the number of Kraus operators $K^{(N)}_{\varphi,k}$ will typically grow  exponentially with $N$. In step (i) we performed the partial trace over spaces $\mathcal{A}_N$ and $\mathcal{K}_N$ and introduced $\tilde{P}=\t{Tr}_{\mathcal{A}_N} {P}$, $\tilde{\Omega}(h) = \t{Tr}_{\mathcal{K}_N} \Omega(h) \in  \mathcal{L}(\mathcal{H}_1\otimes \mathcal{K}_1 \otimes \dots \mathcal{H}_{N-1} \otimes \mathcal{K}_{N-1} \otimes \mathcal{H_N})$, while 
in (ii) we again used the minimax theorem, as optimization over both $h$ and $\tilde{P}$ is over convex spaces and the function itself is convex in $h$ and concave (linear) in $\tilde{P}$. Note that in the special case $N=1$, $\tilde{\Omega}(h) = \alpha(h)$, $\tilde{P}=\rho_{\mathcal{H}}$ and we recover the formula from \eqref{eq:opnorm}. Unlike in \eqref{eq:opnorm}, however, we cannot replace the maximization over the comb by the operator norm due to nontrivial constraints on $\tilde{P}$. 
Nevertheless, if maximization over $\tilde{P}$ is replaced by an appropriate minimization of the dual problem, one may in the end write the double minimization as a single semi-definite programme in a form resembling that of the single channel optimization \eqref{eq:sdp} \cite{Liu2023}:
\begin{align}
\label{eq:sdpcomb}
&F_Q(\Lambda_\varphi^{(N)}) = 4 \min_{\lambda, h, Q^{(k)}} \lambda,  \ \t{subject to } A \succeq 0, \\
\nonumber
&\underset{2\leq k \leq N-1}{\forall}\t{Tr}_{\mathcal{K}_{k}}Q^{(k)} = \openone_{\mathcal{H}_k} \otimes Q^{(k-1)}, \t{Tr}_{\mathcal{K}_1} Q^{(1)}= \openone_{\mathcal{H}_1}
\end{align}
where 
\begin{equation}
\label{eq:matAN}
A = \left( \begin{array}{c|ccc}
        \openone_{\mathcal{H}_{N}} \otimes Q^{(N-1)}& \ket{\dot{\tilde{K}}_{1,1}(h)}  & \hdots &  \ket{\dot{\tilde{K}}_{r,d}(h)} \\ \hline
        \bra{\dot{\tilde{K}}_{1,1}(h)}&  &  &  	 \\
        \vdots&  & \lambda \mathbb{1}_{d r}  &    \\ 
       \bra{\dot{\tilde{K}}_{r,d}(h)} &  &  &            
 \end{array}\right).
\end{equation}
In the above expression, $d = \t{dim}(\mathcal{K}_N)$, 
$\ket{\dot{\tilde{K}}_{i,k}(h)} = {_{\mathcal{K}_N}}\!\!\braket{i|\dot{K}_{\varphi,k}^{(N)}(h)} \in  \mathcal{K}_1 \otimes \mathcal{H}_1 \otimes \dots \mathcal{K}_{N-1} \otimes \mathcal{H}_{N-1} \otimes \mathcal{H}_N$, while dual problem variables $Q^{(k)} \in \mathcal{L}(\mathcal{K}_1 \otimes \mathcal{H}_1 \otimes \dots \otimes \mathcal{K}_k \otimes \mathcal{H}_k)$ are subject to the same quantum-comb constraints, apart from the positivity requirement. 

The above optimization problem will correctly yield the optimal QFI $F_Q^{\diamond}$ of the channel and the corresponding $h^{\diamond}$ matrix. If, however, one wants to identify the corresponding quantum comb ${P}^{\diamond}$ that represents the optimal strategy, one needs to step back to the original primal problem formulation, and find $\tilde{P}$ satisfying the quantum comb constraints such that \cite{Liu2023}: 
\begin{equation}
4 \t{Tr} \left[\tilde{\Omega}(h^{\diamond})  {\tilde{P}}\right] = F_Q^{\diamond}, 
\quad  \nabla_{h=h^{\diamond}} \t{Tr}\left[\tilde{\Omega}(h) \tilde{P} \right] =0.
\end{equation}
Similarly as in \eqref{eq:saddle} the second condition is necessary to make sure the solution is at the saddle point and the resulting $\tilde{P}^{\diamond}$ corresponds indeed to the optimal protocol. Note that in the original paper \cite{Altherr2021} this second condition was not included and hence the procedure described there might not lead to the actual optimal protocol---it has only been remedied in the follow-up paper \cite{Liu2023}.

\subsection{Iterative see-saw (ISS) approach}
\label{sec:multiple_iss}
As a completely new result, we demonstrate how the ISS optimization can be generalized to the multiple coherent uses regime---it turns out to be more straightforward than generalization of the MOP method. The generalization of the `pre-QFI' function from \eqref{eq:frl}, to be maximized, now takes the form: 
\begin{equation}
\label{eq:qfipl}
F^{(N)}({P},L) = 2\t{Tr}\left(\dot{\rho}^{(N)}_\varphi L \right)-\t{Tr}\left(\rho^{(N)}_\varphi L^2\right),
\end{equation}
where $\rho_{\varphi}^{(N)} = {\varLambda}_\varphi^{(N)} \star {P}$.
The optimal QFI is again obtained as a result of double maximization:
\begin{equation}
F_Q(\Lambda^{(N)}_\varphi) = \max\limits_{{P},L} F^{(N)}({P},L),
\end{equation}
with constraints on ${P}$ to be a quantum comb as given in \eqref{eq:combcons}. 
We start the iteration procedure with some randomly chosen initial ${P}^{[0]}$ and find the corresponding $L^{[0]}$ maximizing  $F^{(N)}({P}^{[0]},L)$. This step is identical to the one in the single channel estimation approach and amounts to finding the SLD for the state $\rho_\varphi^{(N)}$. Then fixing $L^{[0]}$ we identify the optimal $P^{[1]}$ and so on.  This second step, while slightly more complex than in the single channel case \eqref{eq:secondstep}, can nevertheless again be written as a relatively simple SDP programme:
\begin{align}
&\max\limits_{{P} }F^{(N)}({P},L) = \max\limits_{{P}}\t{Tr}\left({P} M^{(N)}\right), 
\nonumber\\
&M^{(N)} =  2 L \star  \left(\dot{{\Lambda}}_{\varphi}^{(N)}\right)^T - L^2 \star \left({\Lambda}_{\varphi}^{(N)}\right)^T,
\end{align}
where we used straightforward properties of the link product operation to rewrite \eqref{eq:qfipl} in the above form. This is a linear optimization problem in ${P}$ with convex constraints \eqref{eq:combcons}, hence an SDP.
In each step of the iteration $F^{(N)}$ will not decrease, and we terminate the procedure when it converges up to the desired accuracy.  

Similarly as in the optimization over purifications method, this procedure is efficient provided the dimensions of spaces on which channel $\Lambda_\varphi^{(N)}$ acts are reasonable---for multiple uses of a single qubit channel this usually means $N \leq 5$.   
One of advantages of the ISS approach over the MOP is that we can control the size of the final ancillary space $\mathcal{A}_N$ by restricting optimization over $P$ to combs with a fixed dimension of $\mathcal{A}_N$ (but without the  control of the size of ancillary systems required for the inner action of the comb itself). The main advantage, however, is that the approach can be naturally adapted to the tensor network formalism, as described in Sec.~\ref{sec:tensor}  and in many cases allows to go around the curse of dimensionality problem and study optimal metrological protocols in the limit of large number of coherent channel uses. In this approach it will also be possible to control the size of ancilla at every step of the protocol.

\subsection{Decomposition of a quantum comb into elementary operations}
\label{sec:decomposition_into_isometries}
Given the CJ operator  ${P}$ of the optimal quantum comb, obtained from either the MOP or ISS method, it may be non-trivial to obtain a deeper insight into the character of the metrological protocol it actually represents and how to implement it in practice using a set of simple quantum gates. For this purpose procedures of decomposition of a quantum comb into a sequence of unitary operations (isometries) has been proposed in \cite{Chiribella2009, bisio_minimal_2011}. 
A quantum comb $\t{P}$ can be always rewritten as a concatenation of isometries $\{V^{(k)}\}_{k=1}^{N}$ corresponding to the comb's consecutive teeth, s.t. $V^{(k)}: \mathcal{K}_{k-1} \otimes \mathcal{A}_{k-1} \rightarrow \mathcal{H}_k \otimes \mathcal{A}_k$ (where $\mathcal{A}_{k-1}, \mathcal{A}_{k}$ are the ancillae on the input and output of the $k$-th tooth, respectively and $\mathcal{K}_0, \mathcal{A}_0$ are trivially $\mathbb{C}$). The input and output spaces of the isometries can schematically be pictured as follows:
\begin{equation}
\label{eq:isometry_decomposition}
    \includegraphics[width=0.85\columnwidth]{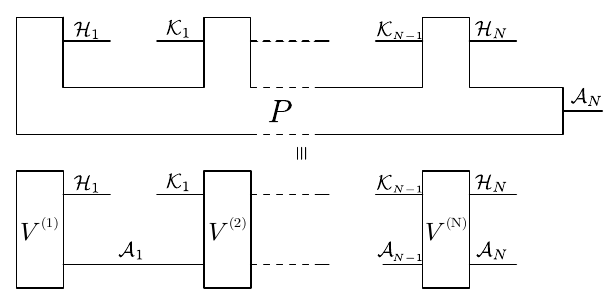}.
\end{equation}
The details on how to find the corresponding isometries are presented in Appendix~\ref{app:decomposition}.

However, constructing these isometries in general will require ancillary systems of dimensions that may grow exponentially with $k$. The minimal possible dimension of ancilla $\mathcal{A}_k$ necessary to represent $V^{(k)}$ is $\mathrm{dim}(\mathcal{A}_k) = \mathrm{rank}(P^{(k)})$ \cite{bisio_minimal_2011}. Rank of $P^{(k)}$ may be as large as the dimension of the space on which the operator acts, which in this case is the product of dimensions of all its input and output spaces $\mathcal{K}_{i-1}, \mathcal{H}_i$, up to $i=k$. From our numerical experience it appears that in typical metrological scenarios, the isometries corresponding to optimal metrological strategies in noisy models obtained from comb optimization methods described above are very complex, and the size of ancillary systems indeed tends to grow significantly with $k$---see the discussion of an example in Sec.~\ref{sec:parallel_damping}. 

Another challenge is that even if the optimal strategy may not be that complex, it may be returned by the optimization procedure in a basis which is not easy to interpret.
In principle one could use automated methods, implemented in e.g. BQSKit \cite{BQskit} , UniversalQCompiler \cite{iten_introduction_2021} of decomposition of isometries into gates from standard gate sets to obtain a practical implementation of the optimal protocols. This approach to describing optimal metrological protocols was taken e.g. in Ref.  \cite{Liu2023}. However, the challenge of interpreting those schemes still remains. 

These difficulties in obtaining a simple and intuitive structure of the optimal protocols are related to substantial freedom when identifying optimal quantum combs. The optimal comb found is typically not unique and its form rarely allows for a direct intuitive understanding of the essence of the protocol. This is another strong argument (apart from the curse of dimensionality issue) in favour of the tensor network approach presented in the next section, where we are able to control the size of the ancillary systems at each `tooth' of the comb, and have means to force the optimization procedure, tooth by tooth, to yield optimal strategies in a form we are able to interpret.
Note that by nature of the approach, this is not possible in the MOP-type methods.

\section{Tensor network approach to identify the optimal adaptive metrological protocols}
\label{sec:tensor}

The main obstacle in finding the optimal adaptive protocols  is the exponential growth of complexity of algorithms presented in Section \ref{sec:multiple} with  increasing $N$. This is related to the size of the CJ operator $P$ representing the estimation strategy comb. Even in the simplest case of qubit channels $\Lambda_\varphi$ and no output ancilla,  $ P$ acts on a Hilbert space of total dimension $2^{2N-1}$. This limits the applicability of introduced methods on present-day personal computers to the cases where $N \le 5$---for larger $N$, SDP memory and time requirements are hard to meet.

Moreover, in the approaches presented so far, the internal structure of $P$ cannot be controlled---in particular, we cannot limit the size of the ancillary system required to implement each tooth of the optimal comb, we can only control the last ancilla $\mathcal{A}_N$. Consequently, the strategies obtained may be
very complex even for small $N$.

To overcome both of these problems, we decompose $P$ into teeth $P_1, P_2, ..., P_N$, representing simpler quantum channels whose concatenation leads to $P$. Formally,
\begin{equation}
\label{eq:P_decompose}
    P = P_1 \star P_2 \star ... \star P_N,
\end{equation}
where $P_1 \in \mathcal{L}( \mathcal{H}_1 \otimes \mathcal{A}_1)$, $P_k \in \mathcal{L} (\mathcal{A}_{k-1} \otimes \mathcal{K}_{k-1} \otimes \mathcal{H}_k \otimes \mathcal{A}_k)$ for $2 \le k \le N$---when the link product between $P_k$ and $P_{k+1}$ is performed the common subspace is $\mathcal{A}_k$. Importantly, $P_i$ can be arbitrary CJ matrices, not necesarily isometries, contrary to the comb decomposition described in Section \ref{sec:decomposition_into_isometries}. This allows for effective optimization over each comb tooth $P_i$, since CJ matrices, unlike isometries, form a convex set. This also changes the meaning of ancillary spaces $\mathcal{A}$---in Section \ref{sec:decomposition_into_isometries} they played dual role of a comb memory and purification, whereas in \eqref{eq:P_decompose} we do not need to purify $P_i$, so ancillary system is only used as a quantum memory. 

The link product is linear in both arguments, and can be represented as contraction of indices between two tensors (see appendix \ref{app:ten_net}  for short introduction to tensor networks formalism and technical details). Consequently, we can write down the RHS of \eqref{eq:P_decompose} as a tensor network, whose nodes (rectangles) represent quantum channels,  open links represent subspaces on which $P$ acts and closed links represent link products:
\begin{equation}
\label{eq:Ptensor}
    \includegraphics[width=1\columnwidth]{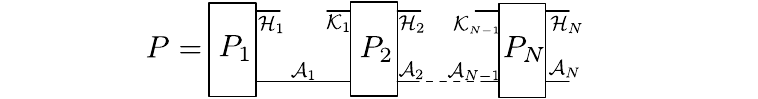}.
\end{equation}
Notice that with this graphic notation it is clear which subspaces must be contracted while performing the link product. 

When output and input spaces of probed channels are all of the same dimension ($\textrm{dim} (\mathcal{H}_k) = \textrm{dim} (\mathcal{K}_k) = d_\mathcal{H}$), and the size of ancilla is fixed during the whole protocol ($\textrm{dim} (\mathcal{A}_k) = d_\mathcal{A}$), then  $d_\mathcal{H}^{4N-2} d_\mathcal{A}^2$ complex variables are required to store $P$ in the memory. We can substantially compress the information about $P$ by storing $P_1, P_2, ..., P_N$ separately, which  requires only  $d_\mathcal{H}^2 d_\mathcal{A}^2 + (N-1) d_\mathcal{H}^4 
d_\mathcal{A}^4$ variables. The latter approach is much more effective for fixed $d_\mathcal{A}$ and growing $N$---the used memory scales linearly, not exponentially with $N$.  To take advantage of this, it is crucial to design an optimization algorithm that operates on $P_1, P_2,...,P_N$ separately, and does not need to refer to the whole $P$. This approach is completely different than the one from Section \ref{sec:decomposition_into_isometries}---instead of finding optimal $P$ and decomposing it, we use ansatz \eqref{eq:P_decompose} from the very beginning to overcome the curse of dimensionality.

The compression of $P$ is only possible for combs which can be written as \eqref{eq:P_decompose} with small $d_\mathcal{A}$.  In general, the size of ancillary system required to simulate all possible combs grows exponentially with $N$, and so does the size of CJ operators $P_k$. However,  it makes sense to search only through strategies $P$ with limited $d_\mathcal{A}$ because such strategies are usually substantially easier to implement in practice. One may also optimize the QFI over $P$ with growing $d_\mathcal{A}$---when the figure of merit no longer increases with increasing $d_{\mathcal{A}}$ it strongly suggests that the optimal strategy $P$ has been  found. 

The analogous idea is used in other tensor networks approximations, in particular in the construction of matrix product state (MPS) representation of entangled states of $N$ particles \cite{Fannes1992, Schollwock2011}. The size of a density matrix of such a system grows exponentially with $N$. However, states that are weakly entangled can be efficiently represented as a tensor network (MPS). Then, the memory required to store the information about a state grows linearly with $N$, but also depends on a so-called bond dimension $d$. The more entangled the state is, the larger $d$ is required. In our case, $d_\mathcal{A}$ plays a role of a bond dimension, with a clear physical interpretation of the size of available ancillary system.

The CJ operator $\varLambda_\varphi^{(N)}$, containing  information about the estimated signal and (possibly correlated) noise, can be also represented as a tensor network. Using the introduced graphical notation:
\begin{equation}
\label{eq:Ltensor}
\includegraphics[width=1.\columnwidth]{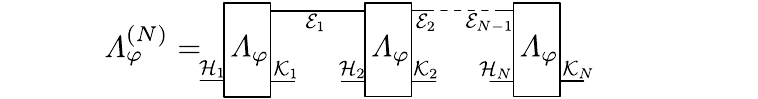},
\end{equation}
where $\mathcal{E}_k$ represents the environment space after the $k$-th use of the channel to be estimated. When the signal and the noise are not correlated, then $ \varLambda_\varphi^{(N)} = \varLambda_\varphi^{\otimes N}$, the state of environment does not affect the action of the subsequent channel, and consequently   $\mathcal{E}_k$ links may be ignored---the tensor network representing $\varLambda_\varphi^{\otimes N}$ is a trivial network without connections. Different types of correlations can be simulated using non-trivial action of channels on $\mathcal{E}_k$---see Section \ref{sec:corr_deph} for an example involving a correlated dephasing noise model. The more complicated and long-range correlations are, the larger the dimension of $\mathcal{E}_k$ required  to simulate them. Again, to simulate all possible signal combs $\Lambda_\varphi^{(N)}$, one would need $\textrm{dim} (\mathcal{E}_k)$ that grows exponentially with $N$. Luckily, for many typical correlation models, the required $\textrm{dim} (\mathcal{E}_k)$ does not depend on $N$ at all.

The problem of identifying the optimal QFI for a sequence of (possibly correlated) channels $\Lambda_\varphi$, with limited size of ancillary system $d_\mathcal{A}$, can be written 
using ISS approach as
\begin{equation}
\label{eq:QFIdA}
    F_Q^{(d_\mathcal{A})}(\Lambda_\varphi^{(N)}) = \max_{P_1,P_2,...,P_N,L} F^{(N)}(P,L)  ,
\end{equation}
 where $P$ is given by \eqref{eq:P_decompose} with links $\mathcal{A}$ of dimensions $d_\mathcal{A}$ and $F^{(N)} (P,L)$ is defined as in \eqref{eq:qfipl}. Obviously, $F_Q^{(d_\mathcal{A})}(\Lambda_\varphi^{(N)}) \le F_Q(\Lambda_\varphi^{(N)})$, and $F_Q^{(d_\mathcal{A})}(\Lambda_\varphi^{(N)}) = F_Q(\Lambda_\varphi^{(N)})$ for sufficiently large $d_\mathcal{A}$.
 
 To perform the maximization \eqref{eq:QFIdA} numerically, we proceed as follows. Initially, $P_1, P_2,...,P_N$ are random CJ operators and $L$ is a random hermitian matrix. Then, we maximize the figure of merit over $P_1$, fixing $P_2,...,P_N,L$. In the next step, we maximize over $P_2$, then over $P_3$, etc. In the final step, the maximization over $L$ is performed. The entire procedure is then repeated until convergence is achieved. Notice, that compared to the algorithm from Section \ref{sec:multiple_iss}, we need to perform $N+1$ maximization steps in one iteration instead of just $2$ steps. However, the computational complexity of each step does not scale with $N$. It is also crucial that each maximization step is SDP (as we show below).

 The  network representing $F^{(N)}(P,L)$ can be depicted as 
 \begin{equation}
 \label{eq:tensor_F}
\includegraphics[width=1.\columnwidth]{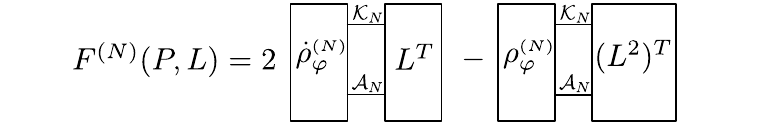},
 \end{equation}
 where we used the identity $A \star B = \t{Tr} (A B^T)$ valid for matrices $A,B$ acting on the same Hilbert space. The  matrices $\rho_\varphi^{(N)}$,  $\dot \rho_\varphi^{(N)}, L$ are all of size $d_\mathcal{H} d_\mathcal{A} \times d_\mathcal{H} d_\mathcal{A}$, which does not depend on $N$. Hence those matrices can be easily stored in memory.
Given  $\rho_\varphi^{(N)}$,  $\dot \rho_\varphi^{(N)}$,  the optimization over $L$ can be performed in the same way as it was done in Sections \ref{sec:single_iss}, \ref{sec:multiple_iss}. The nontrivial part is the efficient computation of matrices $\rho_\varphi^{(N)}$, $\dot \rho_\varphi^{(N)}$. This can be done by contracting the following networks:
\begin{equation}
\label{eq:tensor_rho}
    \includegraphics[width=1.\columnwidth]{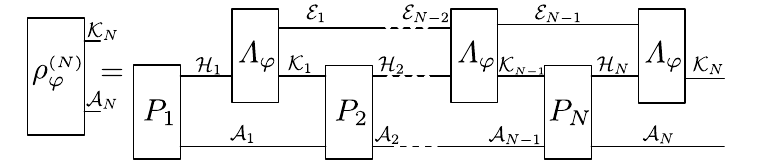},
\end{equation}

\begin{equation}
\label{eq:tensor_drho}
    \includegraphics[width=1.\columnwidth]{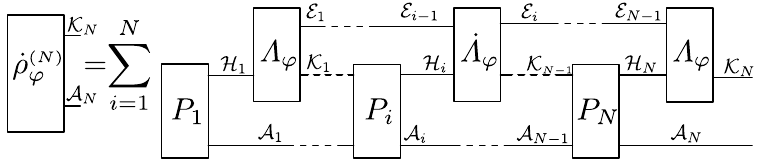}.
\end{equation}

Notice, that $\dot \rho_\varphi^{(N)}$ is represented as a sum of $N$ elements---in the $i$-th element the derivative acts on CJ operator of the $i$-th probed channel $\varLambda_\varphi$.

   The key property of tensor networks
   is the freedom of choice of indices contraction order.
   Even though $\rho_\varphi^{(N)}$ is constructed by contraction of networks representing $P$ and $\varLambda_\varphi^{(N)}$, it is not necessary to compute these constituent networks in order to obtain $\rho_\varphi^{(N)}$ (which would be equivalent to  the procedure from Section\eqref{sec:multiple_iss}). Instead, one should contract indices in the following order: $\mathcal{H}_1, \mathcal{A}_1, \mathcal{K}_1, \mathcal{E}_1, \mathcal{H}_2, \mathcal{A}_2$, ... , $\mathcal{E}_{N-1}, \mathcal{H}_N$. Then,  the maximal size of a tensor we need to process at a time does not depend on $N$.  This allows to compute $\rho_\varphi^{(N)}$ and its derivative efficiently even for large $N$, and then find optimal $L$ in the same way as in previously described ISS procedures.

To optimize over $P_k$, we proceed as follows.  Firstly, we represent the figure of merit using \eqref{eq:tensor_F}. Then, we replace  $\rho_\varphi^{(N)}$ and $\dot \rho_\varphi^{(N)}$ with  networks \eqref{eq:tensor_rho}, \eqref{eq:tensor_drho}. The figure of merit is then represented as a sum of $N+1$ networks: $N$ from the term  $2 \t{Tr} (\dot \rho_\varphi^{(N)} L)$ and 1 from the term  $ \t{Tr}(\rho_\varphi^{(N)} L^2)$. In each component network, we contract all the indices apart from those corresponding to subspaces linked to $P_k$. Then, we obtain
\begin{equation}
\label{eq:tensor_PiSi}
    \includegraphics[width =  0.95\columnwidth]{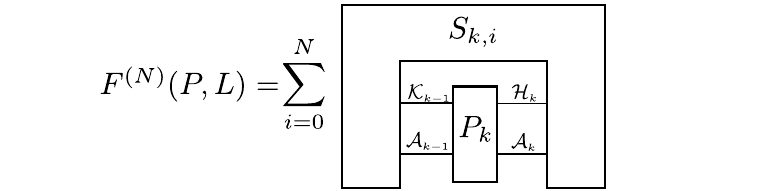},
\end{equation}
where $S_{k,0}$ denotes the contracted network representing term with $ \rho_\varphi^{(N)}$ and $S_{k,i}$ for $i>0$ denotes the $i$-th term from the expansion of the term with $ \dot \rho_\varphi^{(N)}$. CJ operators $S_{k,i}$ act on the Hilbert space of dimension $d_\mathcal{A}^2 d_\mathcal{H}^2$, and their sum
\begin{equation}\label{eq:Sk_sum}
    S_k = \sum_{i=0}^N S_{k,i}
\end{equation}
 can be directly computed, which allows to write the figure of merit as 
  \begin{equation}
      F^{(N)}(P,L) = \text{Tr} (P_k S_k^T).
  \end{equation}
  Interestingly, when proper partial results of computations are saved, then the time complexity of computing $S_1, S_2, ..., S_N$ is $O(N)$, which is also the time complexity of the whole algorithm, see Appendix \ref{app:ten_net_opt}.
The optimization over $P_k$ boils down to the following SDP:
\begin{align}
\label{eq:opt_Pk}
&\max\limits_{P_k }F^{(N)}(P,L) = \max\limits_{P_k}\t{Tr}\left(P_k S_k^T \right),
\nonumber\\
& \t{s. t. } \t{Tr}_{\mathcal{H}_k \otimes \mathcal{A}_k} (P_k) = \mathbb{1}_{\mathcal{K}_{k-1} \otimes \mathcal{A}_{k-1}}, P_k \ge 0.
\end{align}
The first condition for $P_k$ corresponds to the trace preservation of a channel, and should be replaced with $\text{Tr} (P_k) = 1$ for $k=1$ because $P_1$ represents a density matrix of an input state. 

The performance of the iterative optimization is
the most stable when CJ operators $\varLambda_\varphi$ are full-rank. In other cases, we observe that convergence to the optimal value is not always achieved. To overcome this issue, we add an artificial depolarizing noise to each channel $\Lambda_\varphi$. The strength of this noise decays exponentially over the course of running the algorithm, such that its role becomes negligible for final iterations, and the optimization result is unaffected. Due to this improvement, our algorithm is very stable, and converges to the same value with different random inputs.

The algorithm outputs not only the optimal QFI value, $F_Q^{(d_\mathcal{A})}(\Lambda_\varphi^{(N)})$, but also the sequence of  CJ operators $P_k$ corresponding to the optimal estimation protocol. In many cases, this protocol can be further simplified. Let us consider the following transformation of two subsequent comb teeth:
\begin{align}
&\t{P}_k \rightarrow (\mathcal{I}_{\mathcal{H}_k} \otimes \t{U}_{\mathcal{A}_k}) \circ \t{P}_k \\\
&\t{P}_{k+1} \rightarrow \t{P}_{k+1}  \circ (\mathcal{I}_{\mathcal{K}_k} \otimes \t{U}^{\dagger}_{\mathcal{A}_k}),
\end{align}
where $\mathcal{I}$ is an identity channel, $\t{U}$ is an arbitrary unitary channel, $\t{U}^\dagger$ is its inverse and subscript denotes a space on which a channel acts. Channels $\t{U}_{\mathcal{A}_k}$ and $\t{U}^\dagger_{\mathcal{A}_k}$ become identity when the link product $P_k \star P_{k+1}$ is performed---consequently, the described transformation does not change $P$. The proper choice of $\t{U}_{\mathcal{A}_k}$ may simplify the teeth $P_1, P_2, ..., P_k$ of the optimal strategy---this is an analogue of the `local-gauge' choice in the MPS description \cite{Schollwock2011}.  In our algorithm we can fix the initial teeth in an easy to interpret basis, rerun the optimization over the remaining ones and thus, step by step, limit some of the gauge freedom.

\section{Examples}
To demonstrate the efficiency of the introduced tensor network based approach, we use it to find  the optimal adaptive protocols of estimation of different noisy qubit channels and the corresponding QFI in the limit of large number of channel uses. In our numerical implementation to solve SDP problems we used CVXPY package \cite{diamond2016cvxpy} for modeling and MOSEK solver \cite{mosek}.

The methods we have developed can be applied to channels $\Lambda_\varphi$ with arbitrary parameter dependence. Still, for concreteness we focus on qubit channel estimation models with unitary parameter encoding and parameter independent noise, for which the Kraus operators have the following structure:
\begin{equation}\label{eq:general_kraus_form}
K_{\varphi, k} = U_\varphi K_k, \quad  U_\varphi = e^{-\frac{i}{2}\varphi\sigma_z},
\end{equation}
which means that the angle of rotation of the Bloch vector around $z$ axis is estimated, and signal comes after noise described by Kraus operators $K_k$. This allows us to discuss models that manifest qualitatively different behaviour both in terms of asymptotic QFI scaling, and in terms of the impact of the size of available ancillary system.

\begin{figure}[t]    
    \includegraphics[width=1.0\columnwidth]{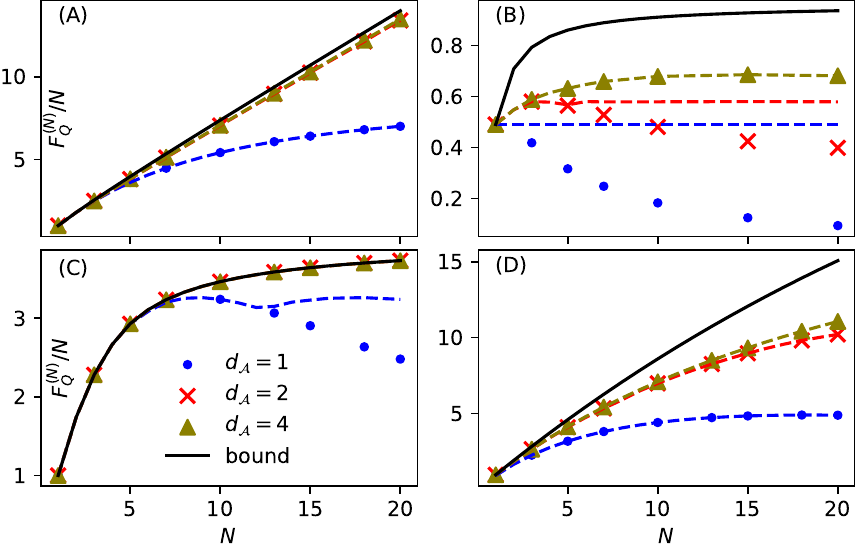}
    \caption{Optimal values of QFI normalised by $N$ for optimal adaptive strategy, given $N$ channel uses, for different dimensions of ancillary system: $d_\mathcal{A}=1$ (blue),  $d_\mathcal{A}=2$ (red) and $d_\mathcal{A}=4$ (yellow). 
    Symbols depict values of QFI in a situation where all the $N$ channels are used in a single adaptive protocol, while dashed lines allow for multiple  repeated experiments with $N$ channels used in total. Solid black lines correspond to the fundamental upper-bound computed using the methods from \cite{Kurdzialek2023}. The four plots correspond to different metrological models: (A) perpendicular dephasing, Eq.~\eqref{eq:per_deph} ($p=0.9$); (B) parallel dephasing, Eq.~\eqref{eq:deph1}($p=0.85$) (C) perpendicular amplitude damping, Eq.~\eqref{eq:per_damp} ($p=0.75$); (D) parallel amplitude damping, Eq.~\eqref{eq:parallel_damping_definition} ($p=0.9$). } 
    \label{fig:examples}
\end{figure}

We will present results on the achievable QFI for optimal adaptive protocols utilizing given size of ancillary systems for four representative types of uncorrelated noise affecting phase estimation: perpendicular dephasing (\ref{sec:perpdeph}), parallel dephasing (\ref{sec:pardeph}), perpendicular amplitude damping (\ref{sec:perp_damp}), and  parallel damping (\ref{sec:parallel_damping})---see Fig.~\ref{fig:examples}. We present the result for number of channel uses up to $N=20$ as in these regime all the relevant qualitative observations can be made, but this is not a fundamental limitation of the method.  As a final example, in \ref{sec:corr_deph} we will present results for a correlated dephasing noise model (for number of channel uses up to $N=50$), which shows how effective the tensor network method is in understanding the potential of correlated noise models, for which the fundamental metrological limitations are not yet fully understood.  

For all the cases studied, we obtain the optimal adaptive QFI as a function of  $N$ for $0$, $1$ and $2$-qubit ancillary systems (this corresponds to $d_\mathcal{A} \in \{1,2,4\}$). We compare our results with fundamental upper-bounds, which were derived in Ref. \cite{Kurdzialek2023}. The bounds are guaranteed  to be saturable when $N \rightarrow \infty$ and ancillary system is large enough \cite{Zhou2020, Kurdzialek2023}. However, for some cases, the gap between the upper bound and the result of ISS optimization 
is very small. It means, that an almost optimal adaptive estimation protocol can be implemented with a small size ancillary system. To guarantee a high precision of the results obtained, we stop the  algorithm only when the relative result increases less than $0.01 \%$ over 5 full iterations (optimization over $P_1$,..., $P_N$, $L$ is counted as one iteration).  We also made sure, that for different, randomly chosen initial guesses $P_1$,..., $P_N$, $L$, the final result is the same up to a numerical error. 

\subsection{Perpendicular dephasing }
\label{sec:perpdeph}
The noise Kraus operators for this model are given by 
\begin{equation}  \label{eq:per_deph}
    K_{1} = \sqrt{p}\,\,\openone, \; K_{2} = \sqrt{1-p}\sigma_x,
\end{equation}
which means that the dephasing acts perpendicularly to the axis of the parameter encoding unitary rotation. This model has long been a paradigmatic example of the potential of application of quantum-error correction inspired protocols to recover the Heisenberg scaling of precision despite presence of noise \cite{Arrad2014, Kessler2014a, Dur2014, Sekatski2016, Demkowicz-Dobrzanski2017, Zhou2017}. 

In case when signal comes before a noise, that is when $K_{\varphi, k}=K_k U_\varphi$ instead of $K_{\varphi, k}=U_\varphi K_k$ \eqref{eq:general_kraus_form}, the impact of noise can be completely eradicated with only one ancillary qubit \cite{Arrad2014, Kessler2014a, Dur2014, Sekatski2016, Demkowicz-Dobrzanski2017, Zhou2017}.

It is also known that when signal comes after the noise, which is the case we consider in this paper, the Heisenberg scaling may also be preserved, yet with a reduced coefficient \cite{Kurdzialek2023}. 
In Ref. \cite{Kurdzialek2023} the optimal protocol in case of $N=2$ uses of channel has also been explicitly constructed, which required no ancillary system at all. It was not clear, however, what is the structure of the optimal protocol for larger $N$ and in particular what size of ancillary system is required to reach the optimal performance.  

In Fig.~\ref{fig:examples}A we see that the ancillary system is indeed needed in order to preserve the character of the Heisenberg scaling, and that a single qubit ancilla already provides an almost optimal performance---the results fall very close to the fundamental bound, and the numerical improvements thanks to the addition of second ancillary qubit are marginal.

The exemplary almost optimal protocol for $N=3$ that utilizes only a single ancillary qubit, that we were able to extract from the obtained numerical results, is described in Appendix~\ref{app:perp_deph_details}. 



\subsection{Parallel dephasing}
\label{sec:pardeph}
Parallel dephasing is one of the most commonly considered decoherence models in quantum metrology, as it represents the typical situation where the coherences required to sense the parameter of interest are being reduced by decoherence processes. The corresponding noise Kraus operators are
\begin{equation}  
\label{eq:deph1}
    K_{1} = \sqrt{p}\,\,\openone, \; K_{2} = \sqrt{1-p}\sigma_z,
\end{equation}
which means that the dephasing is defined with respect to the same axis as the parameter encoding rotation.  

It is well known that this model does not admit asymptotic Heisenberg scaling,
and the quantum enhancement amounts to a constant factor improvement, with the asymptotically achievable upper bound on  QFI given as $F^{(N)} \leq N \frac{(p-1/2)^2}{p(1-p)}$       \cite{Escher2011, Demkowicz-Dobrzanski2012, Demkowicz-Dobrzanski2014}. It is also well known, that in the parallel-scheme framework the bound can be asymptotically achieved via a Ramsey interferometry scheme and the use of weakly spin-squeezed states \cite{Orgikh2001, Escher2011, Chabuda2020} or low bond dimension MPS \cite{Jarzyna2013, Chabuda2020}. These protocols may be practical in many-body systems, as the effecitve entanglement required between the particles is weak, but, nevertheless, large number of elementary probes need to be entangled. As such, it is not obvious if the performance of this optimal parallel many-probe protocol can be effectively simulated via adaptive protocols with small ancilla size in the limit of large $N$. 

Fig.~\ref{fig:examples}B illustrates that increasing size of ancilla significantly improves the performance of the protocol. The largest size considered, $d_\mathcal{A}=4$, shows significantly better performance  than the single qubit ancilla case, but clearly, increasing the dimension of ancilla further would allow to approach the bound even closer---this would require a more dedicated numerical effort though, as three qubit ancilla $d_\mathcal{A}=8$ is on the borderline of numerical complexity that a high performance PC is capable of dealing with.
Note that for low dimensional ancilla, at some point QFI starts to drop down, since the decoherence is dominating, and the ancillary systems are not sufficiently large to deal with. In this case it is more advisable to stop the protocol at some $n<N$, and use the remaining resources for a fresh run of the protocol---this strategy is illustrated with dashed lines, and this allows to avoid the drop in per-channel-use performance with increasing $N$. Interestingly, the effect of QFI per channel descrease is barely noticeable already with a two qubit ancilla. 

One might wonder, how relevant is the adaptive aspect of the protocol, and ask how the optimal entanglement based strategy with the same number of available qubits would perform. 
In Appendix~\ref{app:par_deph_adapvspara} we present numerical results, that confirm that the adaptive strategy with one probe and two ancillary qubits (three qubits total) outperforms the optimal strategy utilizing three entangled qubits. This remains true even if we allow for sequential channel probing (but not full adaptiveness, since intermediate controls are not allowed) in the latter case.  This shows that the adaptive strategy found is not just the imitation of an entanglement based strategy. Moreover, it clearly demonstrates that quantum adaptiveness is more than entanglement $+$ sequentiality.

\subsection{Perpendicular amplitude damping}
\label{sec:perp_damp}

\begin{figure}[h!]
    \includegraphics[width=1.\columnwidth]{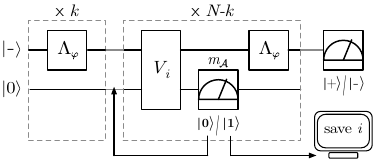}
    \caption{Schematic representation of the optimal metrological protocol for phase estimation in the presence of perpendicular damping noise. Initially, probe and ancilla qubits are prepared in a state $|-\rangle_{\mathcal{H}_1} \otimes |0\rangle_{\mathcal{A}_1}$, and for first $k$ steps, no external control is applied---the probe state is freely evolved through channels $\Lambda_\varphi$. Then, the protocol consists of the interactions $V_i$ between system and ancilla, followed by measurement of ancilla in $|0\rangle/|1\rangle$ basis. If the result of this measurement is $m_\mathcal{A} = 0$ , the protocol is continued, if $m_\mathcal{A} = 1$, the protocol is terminated, and the number of step $i$ in which it happened is saved. After $N$ steps the system is measured in $|+\rangle/|-\rangle$ basis (provided that the protocol was not terminated earlier).} 
    \label{fig:perp_damping_strategy}
\end{figure}

Let us now consider a perpedicular damping model, where
the corresponding Kraus operators read
\begin{equation}\label{eq:per_damp}
    K_1 = \ket{-}\bra{-} + \sqrt{p} \ket{+}\bra{+}, \; K_2 = \sqrt{1-p} \ket{-}\bra{+},
\end{equation}
where $\ket{\pm} = (\ket{0} \pm \ket{1})/\sqrt{2}$ are the eigenvectors of $\sigma_x$. This is variation on the standard amplitude damping model, where the damping axis is perpendicular to the phase encoding rotation axis. 

Despite perpendicular character of the noise, this model, unlike the perpendicular dephasing one, does not admit asymptotic Heisenberg scaling \cite{Kurdzialek2023}. The analysis of this model leads to a particularly interesting conclusions, as
the adaptive upper-bounds for QFI derived in  Ref. \cite{Kurdzialek2023} are saturable for all values of $p$ and $N$. Previously, this fact  was only demonstrated for $N \le 4$. Using the tensor network approach, we show numerically, that the bound is also saturable for $N$ up to 50. Moreover, one qubit ancilla ($d_\mathcal{A} = 2$) is enough to saturate the bound, see Fig.~ \ref{fig:examples}C. Inspired by the structure of the numerical solution, we have found a simple intuitively appealing analytical form of the optimal protocol for all values of $p$ and $N$.

Let us consider the local parameter estimation around the value $\varphi = 0$  (for other values of $\varphi$ one needs to adjust the protocol by proper rotation of a probe qubit). 
Initially, the probe ($\mathcal{H}_1$) and ancilla ($\mathcal{A}_1$) qubits are prepared in a product state $\ket{-}_{\mathcal{H}_1} \otimes \ket{0}_{\mathcal{A}_1}$---notice, that this state is not affected by damping noise. The protocol involves entangling operations $V_i$ applied between the $i$-th and $i+1$-th use of the channel  $\Lambda_\varphi$ followed by the readout of the ancillary qubit. The action of $V_i$ reads:
\begin{align}
\label{eq:Vcontrol1}
    V_i \ket{-} \otimes \ket{0} &= \ket{-} \otimes \ket{0}, \\ 
    \label{eq:Vcontrol2}
     V_i \ket{+} \otimes \ket{0} &= t_i \ket{+} \otimes \ket{0} + \sqrt{1-t_i^2} \ket{-} \otimes \ket{1},
\end{align}
where $t_i \in [0,1]$ describes the coupling strength between the probe and the ancilla. For $t_i = 0$ the coupling is the strongest, and probe is measured in $\ket{\pm}$ basis, with readout saved on ancillary qubit; for $t_i = 1$, there is no coupling, the input state remains unaffected. All intermediate values describe a weak measurement, when information accumulated in a probe is partially transformed to ancilla. 

After the action of $V_i$, the ancillary qubit is measured in the basis $\{\ket{0}, \ket{1}\}$. When the result is $\ket{1}$, then the probe no longer carries any information about $\varphi$,  the whole protocol is restarted, and the number of a step in which it happened is saved (therefore, there is no need to specify how $V_i$ acts on states of the form $\ket{\psi}_\mathcal{H} \otimes \ket{1}_\mathcal{A}$). When the result is $\ket{0}$, the protocol is uninterrupted. Provided  $\ket{0}$ is measured in all $N$ subsequent coherent uses of $\Lambda_\varphi$, then the final probe state is measured in the basis  ($\ket{\pm}$). The total classical FI achieved using this scheme depends on coupling parameters $t_i$. When the  values of $t_i$ are chosen optimally the bound derived in \cite{Kurdzialek2023} is saturated for all $N$.

In particular we observe, that when the damping noise is weak enough, then first $k$ optimal coupling parameters are $t_i =1$---that means, that one should initially let the probe freely evolve through channels $\Lambda_\varphi$. After this stage, interactions $V_i$ effectively keep the probe state in some optimal, fixed point, in which the effect of  damping noise is not too large, and at the same time, the measurements performed on ancillary qubit are as informative as possible. See Fig. \ref{fig:perp_damping_strategy} for the sketch of the described strategy and Appendix~\ref{app:perp_damp_details} for the derivation and more details.

Apart from one ancillary qubit, the described protocol requires  classical memory to store information about the step number in which $\ket{1}$ was measured. The protocol found by our algorithm does not need any extra memory by construction---this can be understood in a way that information instead of being extracted by the measurement is constantly being feed-forward in the structure of the state. However, we
decided to present the measurement based protocol thanks to more intuitively appealing nature. Notice, that it is not possible to saturate the bound without ancilla (see Fig.~\ref{fig:examples}C, points corresponding to $d_\mathcal{A}=1$).

\subsection{Parallel amplitude damping}
\label{sec:parallel_damping}
For completness of discussion of uncorrelated noise models, we consider the standard amplitude damping model representing e.g. spontaneous emission, where decoherence axis is parallel to the phase encoding rotation axis. The Kraus operators for this model are
\begin{equation}
\label{eq:parallel_damping_definition}
    K_1 = \ket{0}\bra{0} + \sqrt{p} \ket{1}\bra{1}, \; K_2 = \sqrt{1-p} \ket{0}\bra{1}.
\end{equation}
In this model, again, the Heisenberg scaling is asymptotically unattainable \cite{Demkowicz-Dobrzanski2012, Kurdzialek2023}, as can be also seen in Fig.~\ref{fig:examples}D where the curve representing the bound on $F^{(N)}/N$ (black solid) reveals the asymptotic convergence to a constant for large $N$.

 Similarly to the parallel dephasing model, we see that the performance of the protocol improves significantly with increasing size of the ancillary system, and the $2$-qubit ancilla results could be further improved by increasing $d_\mathcal{A}$.  It can also be seen that while a strategy with 1 ancilla qubit offers a significant advantage over the one without ancilla for all $N>1$, the difference between strategies with 1 or 2 ancilla qubits are negligible for small $N$. 
 
\begin{figure}[h!]
    \includegraphics[width=0.85\columnwidth]{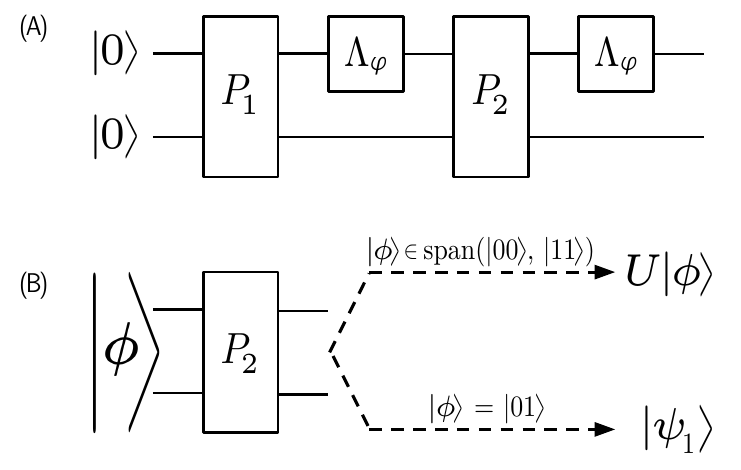}
    \caption{Optimal strategy for parallel damping noise model with two uses of channel $\Lambda_\varphi$, when ancilla is restricted to 1 qubit. a) Teeth $P_1, P_2$ of the strategy act before each channel use. First tooth of strategy, $P_1$, prepares an entangled state. After channel $\Lambda_\varphi$ acts for the first time, the 2-qubit state is $|\phi\rangle$. b) Second tooth, $P_2$, acts on $|\phi\rangle$ with a unitary $U$ acting only on subspace spanned by $|00\rangle, |11\rangle$ or prepares a new entangled state $|\psi_1\rangle$ if the input state to $P_2$ was $|01\rangle$.}
    \label{fig:par_damping_1qubit_ancilla}
\end{figure}

We use this example to demonstrate that the tensor network approach provides benefits also in the low-$N$ regime, by allowing for identifying a simpler structure of optimal protocols than the standard full-comb optimization procedures. More concretely, we focus on the basic $N=2$ uses case, and $p=0.5$, and show a significantly simpler protocol than the one obtained in \cite{Liu2023}, where the MOP optimization, combined with standard procedure of decomposing combs into isometries (discussed in Section \ref{sec:decomposition_into_isometries}) was applied.

Because the difference between QFIs achieved with one- and two-qubit ancillae is small for $N=2$, here we present the strategy with one-qubit ancilla, which is more intuitive and almost optimal (for $N=2$ the QFI changes from $2.174$ to $2.179$ when the second ancillary qubit is added). The 2-qubit ancilla strategy achieves the maximal possible QFI and is described in Appendix~\ref{app:par_damp_details}.

The optimal strategy with one-qubit ancilla is schematically presented in Fig.~\ref{fig:par_damping_1qubit_ancilla}. Its first tooth prepares an entangled state $|\psi_0\rangle$, which is close to the maximally entangled state $\frac{1}{\sqrt{2}}(|00\rangle + |11\rangle)$. The second tooth applies a unitary $U$ on the subspace spanned by $|00\rangle, |11\rangle$ and prepares a different entangled state $|
\psi_{1}\rangle $ for an input state  $|01\rangle$ (notice, that input $|10\rangle$ is forbidden due the noise character). See Appendix~\ref{app:par_damp_details} for more detailed description of this protocol. 

 In \cite{Liu2023} the authors provided a decomposition of the resulting isometries into elementary gates using the package from Ref.~\cite{iten_introduction_2021}. The first tooth of their strategy's comb prepares a pure state. To represent the action of the second tooth, 33 CNOT gates acting between 5 qubits were required.
 As demonstrated above, the second tooth of the strategy which we obtained from our tensor network method, restricting ancilla to one qubit, is significantly simpler and provides an intuitve understanding of the action of the protocol. Even strictly optimal strategy involving two ancillary qubits is much simpler than the one presented in Ref.~\cite{Liu2023}, as we demonstrate in Appendix~\ref{app:par_damp_details}.

\subsection{Correlated parallel dephasing}
\label{sec:corr_deph}
The examples discussed above covered four uncorrelated noise models. For such models, asymptotically tight bounds can be efficiently derived, and serve as a benchmark of actual protocols \cite{Escher2011, Demkowicz-Dobrzanski2012, Demkowicz-Dobrzanski2014, Demkowicz-Dobrzanski2017, Kurdzialek2023}.   

We now move on to show the potential of tensor network methods to deal with correlated noise models. 
In this case, there are no universal methods to derive fundamental bounds, and hence effective numerical methods to identify optimal metrological protocols are even more desired. We will focus on a simple generalization of parallel dephasing model, that would allow us to study the impact of the strength of temporal (anti-)correlations present in the dephasing process. 

Parallel dephasing noise can be interpreted as random rotations of a qubit around the dephasing axis. In the uncorrelated case, these random rotations are assumed to be independent for subsequent interactions of a probe state with the probed channels. In what follows, we will consider a model where these random rotations may be (anti)correlated up to the desired degree. 

The physical context for such a model is a situation where we measure the value of a constant magnetic field with a known direction using a spin $1/2$ probe. The probing spin is also affected by a randomly fluctuating field of a different origin, whose direction is parallel to the field we want to estimate. Moreover, if the time scale of the field fluctuations is comparable or slower than the system probing dynamics, then the random rotations representing dephasing in the subsequently probed channels will be correlated. Therefore, $\Lambda_\varphi^{(N)} \ne \Lambda_\varphi^{\otimes N}$, and the noise is properly described by tensor network with nontrivial Hilbert spaces $\mathcal{E}_i$---see \eqref{eq:Ltensor}.

The one-qubit dephasing of strength $p$, defined in \eqref{eq:deph1}, can be alternatively described using a different set of Kraus operators
\begin{equation}
    \label{deph2}
    K_1 = \frac{1}{\sqrt{2}} U_{+ \epsilon}, \quad K_2 = \frac{1}{\sqrt{2}} U_{- \epsilon},
\end{equation}
where $U_{\pm \epsilon} = e^{\mp \frac{i}{2} \epsilon \sigma_z}$,  $p= \cos^2(\epsilon/2)$. This implies, that the dephasing model can be as well interpreted as resulting from a rotation by angle $\epsilon$ around $z$ axis in random direction ($50 \%$ left, $50\%$ right).

To represent the most basic form of dephasing correlations, we assume that the rotational directions for consecutive dephasing channels are elements of a binary Markov chain given by  
\begin{align}
    \label{eq:pcond1}
    p_{i|i-1}(+ | +) &= p_{i|i-1}(-|-) = \frac{1+C}{2}, \\
    \label{eq:pcond2}
    p_{i|i-1}(+ | -) &= p_{i|i-1}(-|+) = \frac{1-C}{2},
\end{align}
where  $p_{i|i-1}(s_i|s_{i-1})$ for $i \in \{2,3,...,N\}$, is the conditional probability of rotational direction $s_i$ in channel $i$ assuming direction $s_{i-1}$ in channel $i-1$, $C \in [-1,1]$ is a correlation parameter: $C=0$ corresponds to no correlations, $C=1$ means maximal positive correlations (all rotations are in the same direction), and $C=-1$---maximal negative correlations (directions are always different in two neighbor channels).  Furthermore, we assume that directions $+$ and $-$ are equally probable in the first channel: $p_1(+) = p_1(-) = 1/2$.

To model this type of correlations using our tensor networks framework, we consider channels $\Lambda_\varphi$ acting on a two-qubit space: first qubit ($\mathcal{K}$) is the physical probe, and the second one ($\mathcal{E}$) is a classical bit of memory. When this classical bit is in a state $\ket{\pm}$, then the unitary $U_{\pm \epsilon}$ acts on the probe.  After each channel use, the register state is drawn according to the conditional probabilities described in \eqref{eq:pcond1}, \eqref{eq:pcond2}.  The Kraus operators describing the action of a channel on a probe and register are
\begin{align}
&K_1 = \sqrt{\frac{1+C}{2}}U_{+ \epsilon} \otimes \ket{+} \bra{+},~~K_2 = \sqrt{\frac{1-C}{2}}U_{+ \epsilon} \otimes \ket{-} \bra{+} , \\ 
&K_3 = \sqrt{\frac{1-C}{2}}U_{- \epsilon} \otimes \ket{+} \bra{-},~~K_4 =  \sqrt{\frac{1+C}{2}}U_{-\epsilon} \otimes \ket{-} \bra{-}.
\end{align}
The input state of a register of a 1st channel is $\openone /2$ to satisfy the condition $p_1(+) = p_1(-1) = 1/2$. Notice, that the action of a single channel on a probe is equivalent to dephasing of strength $p = \cos^2 \epsilon$ when there is no information about remaining  channels and about classical register state. 

\begin{figure}[th]

\includegraphics[width=0.85\columnwidth]{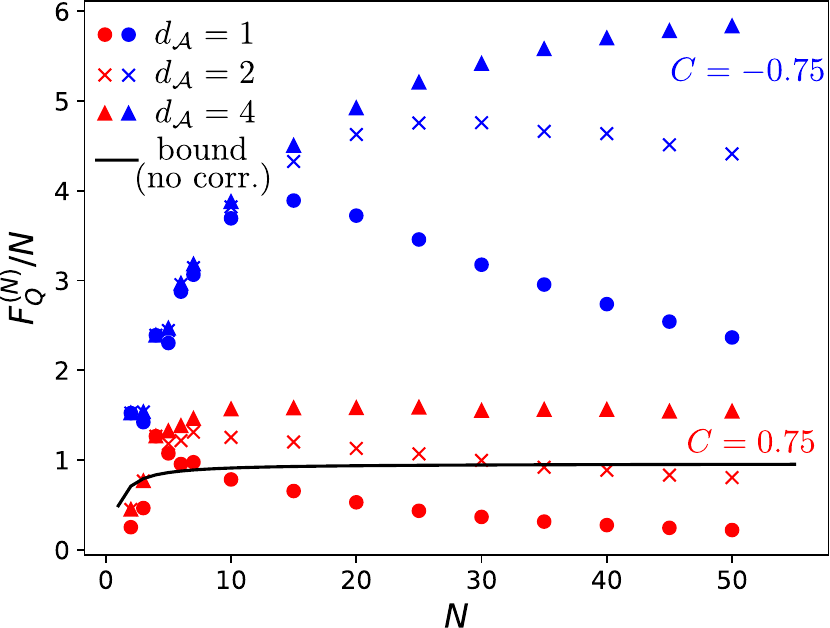}
\caption{The QFI per channel  ($F_Q^{(N)}/N$) as a function of number of channels coherently probed ($N$) in presence of correlated dephasing noise. The data shown were obtained for dephasing strength $p=0.85$, for negative ($C=-0.75$, blue) and positive ($C=0.75$, red) correlations. For comparison, we also show an upper bound for precision for uncorrelated dephasing noise (black line). Interestingly, this bound can be violated with the help of correlations. For small values of $N$, zero and one-qubit ancilla ($d_\mathcal{A} = 1,2$ ) seem to be enough to obtain an optimal precision. However, for larger $N$, the strategy involving two ancillary qubits ($d_\mathcal{A} = 4$) performs substantially better.  Notice, that calculations for $N=50$ would be impossible without tensor network decomposition technique.}
    \label{fig:corr_deph}
    \end{figure}

We performed numerical calculations for negative ($C=-0.75$) and positive ($C = 0.75$) correlations for dephasing strength $p = 0.85$. The results are shown in Fig.~\ref{fig:corr_deph}. 
Note that we also plotted the bound corresponding to the uncorrelated dephasing noise model of the same strength. 
Interestingly, both negative and positive correlations allow to beat the upper bound calculated for uncorrelated case. However, negative correlations allow for significantly larger values of QFI. In fact, we observed that positive correlations may even decrease the QFI for smaller values of $C$. 

The gain in precision related with negative spatial noise correlations fluctuation is a well known phenomena, and has been discussed in parallel quantum metrological schemes \cite{Dorner2012, Jeske_2014, layden2018spatial, Altenburg2017, Chabuda2020}. Note, however, that here for the first time, we provided results of optimal performance of adaptive protocols for \emph{time}-correlated dephasing models in the limit of large $N$ and with arbitrarily tunable correlation parameter.   

The gain thanks to positive correlations may be less intuitive when trying to base it on spatial-correlations analogy. In this case, one would expect that the more collective character of dephasing (positive correlations) leads to even stronger reduction of the achievable QFI---the correlated dephasing \cite{Knysh2014, Macieszczak2014} makes the noise more similar to the signal we are sensing. 
This observation requires a deeper analysis, in order to understand how much of this effect is due to discretization of the phase fluctuations model, and how much is the advantage that appears thanks to the adaptive nature of the protocols considered.

\section{Conclusions and Outlook}
\label{sec:beyondqfi}

The methods presented here allow for the efficient identification of optimal adaptive protocols in the paradigm of multiple coherent channel uses. These results may be seen as complementary to the research focusing on the derivation of fundamental bounds for the performance of adaptive metrological protocols \cite{Demkowicz-Dobrzanski2014, Demkowicz-Dobrzanski2017, Zhou2017,  Kurdzialek2023}. In the latter case, the obtained results are guaranteed to be larger or equal to the largest achievable QFI---if numerical optimization is not exact, the final result can be only too large, never too small. On the contrary, in the ISS optimization, the obtained QFI can never be larger than the optimal one, since we always construct an explicit protocol that allows to achieve the QFI returned by an algorithm. Provided the QFI obtained from the two approaches coincide, we are sure that the protocol we have identified is the optimal one. Interestingly, independently of our paper, a recent study appeared where metrological bounds are discussed taking into account limited size of ancillary systems \cite{Zhou2024}. This approach may be viewed as complementing our results even better, as one may now focus both on protocols and bounds under the same assumptions regarding the limit on the ancillary system size. 

The continuous-time fundamental upper-bound for quantum sensing precision \cite{Wan2022} can be obtained by taking the limit of infinitesimally short time step in a more general discrete bound \cite{Kurdzialek2023}. The similar approach can be used to look for optimal continuous-time quantum control---one can search for optimal discrete-time adaptive strategy with decreasing time step. The only problem is that for a fixed total measurement time, more and more steps are required to make a single step short enough. Therefore, it is impossible to efficiently generalize methods relying on optimization over a full comb to continuous-time regime. However, with our tensor-network based optimization one can deal with much larger numbers of channels---therefore, it is possible to divide the whole time evolution into many small pieces, which can be interrogated with control operations. This allows for a good approximation of continuous control assisted estimation. 

The methods presented here may also be generalized to fit into the Bayesian framework. This in principle poses no conceptual difficulty within the ISS approach, provided the Bayesian cost is quadratic, as the relevant figure of merit has an analogous structure to the QFI, see \eqref{eq:bayesiancost}. The potential practical difficulty may come, however, from a more greedy character of the Bayesian approach when it comes to the dimension of the ancillary systems required---since the classical information is retrieved only in the end of the protocol, and Bayesian approaches typically require larger amount of information to be retrieved in the measurement compared to the QFI based approaches \cite{Hall2012}. Nevertheless, it is worth exploring this direction and compare it with other proposals, where 
methods of identifying optimal adaptive Bayesian protocols in terms of optimization over quantum combs are proposed \cite{Bavaresco2023}.

The other, and probably more promising application is to use the tensor network approach developed here to find the optimal adaptive protocols for channel discrimination problems \cite{Katariya2020a, Bavaresco2021, Katariya2021, bergh2024}.
In this case, even though the problem is also Bayesian at its core, the amount of information gathered over the run of the protocol is limited by the number of alternatives to be discriminated, and protocols with small size ancillary systems should be efficient.
The advantage is the possibility to study the performance of the protocols in the limit of large number of channel uses, a regime out of reach for the all state-of-the art methods \cite{Bavaresco2022}.

Our approach may also be viewed as complementary to tensor network methods developed for identification of optimal multi-partite probe states for parallel sensing schemes \cite{Chabuda2020}. While these approaches are well-suited to study many particle systems (cold atoms, solid state systems, \dots) with simple local entanglement structure, the newly developed approach is perfect to study small scale systems that may be interrogated coherently over many rounds of an experiment (atomic clock systems, trapped ions, spins, NV-centers, \dots). In other words, the present approach addresses the difficulties related with understanding long-time quantum coherence/entanglement potential that may be revealed via specific adaptive protocols, while the former one focuses on spatial aspects of distributed entanglement in multi-partite sensing systems. Interestingly, it is conceivable to combine the two approaches in a unified framework, where both long-range spatial and temporal effects may be satisfactorily analyzed for the sake of identifying optimal metrological protocols. This is an ambitious direction that may be pursued in the future.    

Finally, even though we have focused our research on the standard single parameter quantum metrology problem, it should be possible to generalize the methods presented here to multiparameter models as well. To do this one needs to design an effective ISS procedure involving multiparameter figures of merit such as the exact multiparameter cost or the Holevo bound \cite{DemkowiczDobrzanski2020, Albarelli2019} 
or combine the approach presented with that latest conic-programming based tools for multi-parameter estimation problems proposed in \cite{Hayashi2024}.

We are also convinced that our approach may be fruitfully combined with more general quantum process tomography paradigms, where tensor network structures naturally appear \cite{Pollock2018, White2022}, as well as help in optimizing quantum control operations in non-Markovian models \cite{Butler2024}.  

\emph{Note added}
A week after our manuscript was published, an independent preprint with similar results appeared on arXiv \cite{liu2024efficienttensor}. 

\emph{Acknowledgements.} 
We thank Francesco Albarelli, Jessica Bavaresco, Marco Tulio Quintino, Wojciech G{\'o}recki, David Gross, Kavan Modi, Ingo Roth and Pavel Sekatski for helpful discussions.
This work was supported by the National Science Center (Poland) grant No.2020/37/B/ST2/02134. 

\bibliography{adaptive-tensor.bib}

\clearpage
\newpage

\appendix
\section{Decomposition of a quantum comb into isometries}
\label{app:decomposition}

Here we summarize the procedure, given in \cite{Chiribella2009}, of decomposing a quantum comb $\t{P}$ into a concatenation of isometries (see \eqref{eq:isometry_decomposition} for an illustration) applied to our setup.

Given the CJ operator $P$ of the comb $\t{P}$, we construct the subsequent isometries $V^{(1)}, ..., V^{(N)}$, where $V^{(k)}: \mathcal{K}_{k-1} \otimes \mathcal{A}_{k-1} \rightarrow \mathcal{H}_k \otimes \mathcal{A}_k$. We use the following indices to label the respective basis elements or corresponding Kraus operators:

\begin{tabular}{lcl}
        $i$ & - & basis of $\mathcal{A}_{k}$, $\ i \in [1, \text{rank}(P^{(k)})]$,   \\ 

        $j$ & -  & basis of $\mathcal{A}_{k-1}$,  $\ j \in [1, \text{rank}(P^{(k-1)})]$, \\   

         $m$ & - & basis of $\mathcal{H}_{k}$,  \\

         $n$ & -  & basis of $\mathcal{K}_{k-1}$.    
  \\
  &&
    \end{tabular}



\noindent The procedure is as follows.
\begin{enumerate}
\item Obtain $\forall_{k \in [1, N-1]}$ the CJ operator $P^{(k)}$ of the comb $\t{P}$ up to its $k$-th tooth, using \eqref{eq:combcons}.

\item Starting with $k=1$ for each $k$ construct the canonical representation $\{K^{(k)}_i\}_i$ of the channel $\t{P}^{(k)}$.

\item Starting with $k=2$ (since for $k=1$ the isometry is just the input state), for each $k$ construct two different Kraus representations of the same channel \eqref{eq:decomp_chan1} (see also \eqref{eq:k_prim} and \eqref{eq:k_bis}):
    \begin{equation}
    \label{eqn:reduced_krauses_2}
        \left\{K'^{(k)}_{i, m}\right\}_{i, m} = \left\{\left(\mathbb{1}_{\mathcal{H}_{1, \dots, k-1}} \otimes \langle m | \right)  K_i ^{(k)}\right\}_{i, m},
    \end{equation}
    \begin{equation}
    \label{eqn:extended_krauses_2}
         \left\{K''^{(k)}_{j, n}\right\}_{j, n}= \left\{ K_j ^{(k-1)} \otimes \langle n|\right\}_{j, n}.
    \end{equation}

\item Using the system of equations
\begin{equation}
\label{eq:isometric_relation}
    K'^{(k)}_{i, m} = \sum_{j, n} V^{(k)}_{im, jn} K''^{(k)}_{j, n}
\end{equation}
obtain the matrix elements of the isometry $V^{(k)}$ which connects the two representations of the channel. Finally from:
\begin{equation}
\label{eqn:isometry_final_recipe}
    V^{(k)} = \sum_{i, j, m, n} V^{(k)}_{im, jn} |m \rangle \langle n| \otimes |i\rangle \langle j|
\end{equation}
obtain the isometry. From this construction we have $\mathrm{dim}(\mathcal{A}_k) = \mathrm{rank}(P^{(k)})$.
\end{enumerate}

To see why such $V^{(k)}$ are indeed isometries whose concatenation yields $\t{P}$, let us first recall from \eqref{eq:combcons} the condition for $P$ to be the CJ operator of a quantum comb $\t{P}$:
\begin{align}
\label{eq:appendix_combcons}
&P\geq 0, \ \t{Tr}_{\mathcal{A}_N \otimes \mathcal{H}_N} {P} = {P}^{(N-1)} \otimes \openone_{\mathcal{K}_{N-1}}, \\ 
\nonumber
&\underset{1 <  k<N}{\forall} \t{Tr}_{\mathcal{H}_k} {P}^{(k)} = {P}^{(k-1)} \otimes \openone_{\mathcal{K}_{k-1}}, \ \t{Tr}_{\mathcal{H}_1} {P}^{(1)}=1.
\end{align}
Note that those conditions imply that all $P^{(k)} \in \mathcal{L}(\bigotimes_{i=1}^{k-1}\mathcal{K}_i \otimes \bigotimes_{i=1}^{k}\mathcal{H}_i)$ are positive semidefinite and if we trace out all output spaces they act on ($\mathcal{H}_i: i = 1, ..., k$) we get an identity operator on all input spaces ($\mathcal{K}_i: i = 1, ..., k-1$). Hence each $P^{(k)}$ is a CJ operator of a CPTP map \cite{Chiribella2009}
\begin{equation}
    \mathrm{P}^{(k)} : \mathcal{L}\left( \bigotimes_{i=1}^{k-1}\mathcal{K}_i \right) \rightarrow \mathcal{L}\left(\bigotimes_{i=1}^{k}\mathcal{H}_i\right).
\end{equation}
Through the analysis of the input and output spaces of this map it is easy to see that it is just a quantum channel obtained from comb P by ignoring (tracing out) all spaces related to teeth $k+1, ..., N$ that is $\mathcal{K}_k$, $\mathcal{H}_{k+1}$,  ..., $\mathcal{K}_{N-1}$, $\mathcal{H}_{N}$, $\mathcal{A}_N$. Then the conditions \eqref{eq:appendix_combcons} (apart form positivity) can be reformulated as
\begin{equation}
\label{eq:two_channel_constructions}    \mathrm{Tr}_{\mathcal{H}_{k+1}}\mathrm{P}^{(k+1)}(\rho) = \mathrm{P}^{(k)}(\mathrm{Tr}_{\mathcal{K}_{k}}\rho),
\end{equation}
for any density matrix $\rho$. Naturally, each channel $\t{P}^{(k)}$ can be represented by a set of Kraus operators $\{K^{(k)}_i\}$ and purified to isometry $W^{(k)}$  such that
\begin{equation}\label{eq:kraus_from_iso}
    K^{(k)}_i = \bra{i} W^{(k)},
\end{equation}
where $\ket{i}$ is an o.-n. basis of auxiliary space $\mathcal{A}_k$ used for purification. In the above formula partial inner product was used, that is on the RHS $\langle i|$ formally denotes $\mathbb{1}_{\mathcal{H}_1, ..., k} \otimes \langle i|$. This notation will also be  used throughout this section, when applicable.

Let us for a moment assume that P can be decomposed into isometries $V^{(1)}, ..., V^{(N)}$ and investigate what conditions they should satisfy. First, since $V^{(k)}$ performs the action of the $k$-th tooth, it needs to have $\mathcal{K}_{k-1}$ as one of its input spaces and $\mathcal{H}_k$ as one of its output spaces. Furthermore, among its input spaces we need a space that would connect it with all previous $k-1$ teeth. The natural choice for this space is $\mathcal{A}_{k-1}$, since it can be interpreted as a comb memory after $k-1$ teeth of a comb. Analogously, we need a space that would connect $V^{(k)}$ to the next tooth and the natural choice here is $\mathcal{A}_k$. Thus we are looking for an isometry
\begin{equation}
    V^{(k)}: \mathcal{K}_{k-1} \otimes \mathcal{A}_{k-1} \rightarrow \mathcal{H}_k \otimes \mathcal{A}_k.
\end{equation}
For such a choice we can multiply the purification of $k-1$ teeth ($W^{(k-1)}$) by $V^{(k)}$ and for $V^{(k)}$ to be the $k$-th tooth we want the result of this multiplication to be the purification of $k$ teeth---$W^{(k)}$. In summary, we need $V^{(k)}$ to satisfy:
\begin{equation}
\label{eqn:relation_between_isometries}
    W^{(k)} = V^{(k)}W^{(k-1)}.
\end{equation}

To see that such a matrix indeed exists let us consider two transformations:
\begin{align}\label{eq:decomp_chan1}
    &\mathcal{L}\left( \bigotimes_{i=1}^{k-1} \mathcal{K}_i \right) \ni \rho \mapsto \mathrm{Tr}_{\mathcal{H}_{k}}\mathrm{P}^{(k)}(\rho) \in \mathcal{L}\left( \bigotimes_{i=1}^{k-1} \mathcal{H}_i \right),\\
    &\mathcal{L}\left( \bigotimes_{i=1}^{k-1} \mathcal{K}_i \right) \ni \rho \mapsto \mathrm{P}^{(k-1)}(\mathrm{Tr}_{\mathcal{K}_{k-1}}\rho) \in \mathcal{L}\left( \bigotimes_{i=1}^{k-1} \mathcal{H}_i \right).\label{eq:decomp_chan2}
\end{align}
Clearly both of them are CPTP maps thus they have Kraus representations - here $K'^{(k)}$ and $K''^{(k)}$ respectively. Those representations can be constructed as follows:
\begin{align*}
    \mathrm{Tr}_{\mathcal{H}_{k}}\mathrm{P}^{(k)}(\rho) &= \sum_{m} \bra{m} \mathrm{P}^{(k)}(\rho) \ket{m} =\\
    &= \sum_{i, m} \bra{m} K^{(k)}_i \rho K^{(k)\dagger}_i \ket{m},
\end{align*}
so
\begin{equation}\label{eq:k_prim}
    K'^{(k)}_{i, m} = \bra{m} K^{(k)}_i.
\end{equation}
Analogous reasoning leads to
\begin{equation}\label{eq:k_bis}
    K''^{(k)}_{j, n} = K^{(k-1)}_j \bra{n}.
\end{equation}
By relation \eqref{eq:two_channel_constructions} channels \eqref{eq:decomp_chan1} and \eqref{eq:decomp_chan2} are equal thus $K'^{(k)}$ and $K''^{(k)}$ are two Kraus representations of the same channel. It follows that there exists an isometry $V_{im, jn}$ connecting those two representations \cite{Bengtsson2006}. They are related by \eqref{eq:isometric_relation}.
Finally:
\begin{align*}
    W^{(k)} \overset{\eqref{eq:kraus_from_iso}}{=} \sum_i \ket{i}  K^{(k)}_i \overset{\eqref{eq:k_prim}}{=} \sum_{i, m} \ket{m} \ket{i}  K^{'(k)}_{i, m} =\\
    \overset{\eqref{eq:isometric_relation}}{=} \sum_{i, j, m, n} V_{im, jn}\ket{m} \ket{i}  K^{''(k)}_{j, n} =\\
    \overset{\eqref{eq:k_bis}}{=} \sum_{i, j, m, n} V_{im, jn}\ket{m}\bra{n} \otimes \ket{i}  K^{(k-1)}_{j} =\\
    \overset{\eqref{eq:kraus_from_iso}}{=} \sum_{i, j, m, n} V_{im, jn}\ket{m}\bra{n} \otimes \ket{i}\bra{j} W^{(k-1)} =\\
    = \left( \sum_{i, j, m, n} V_{im, jn}\ket{m}\bra{n} \otimes \ket{i}\bra{j} \right) W^{(k-1)},
\end{align*}
thus the matrix in brackets is an isometry that satisfies \eqref{eqn:relation_between_isometries} and this is indeed the matrix \eqref{eqn:isometry_final_recipe}  which we give in the last step of our procedure.

\section{Tensor network representation of quantum combs}
\label{app:ten_net}
Let us introduce the basic concepts of tensor networks formalism. Any $n$-index tensor $T_{i_1 i_2...i_n}$ can be represented as a rectangle with $n$ legs:
\begin{equation}
    \includegraphics[width=1.\columnwidth]{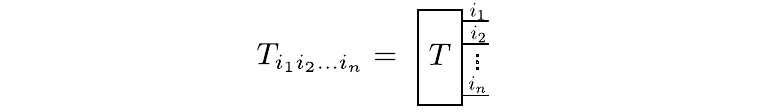}
\end{equation}
All the links in the depicted example come from the right side of a rectangle, generally the side from which a link comes out does not have any mathematical meaning.

Let $T_{i_1 i_2 ... i_n}$ and $W_{j_1 j_2 ... j_m}$ be two tensors, and let us assume that the range of indices $i_k$ and $j_l$ is the same: $i_k,j_l \in \{ 0,1,...,d-1 \}$. Then, a tensor $S$ can be constructed by contraction of indices $i_k$ and $j_l$:
\begin{equation}
\label{index_contraction}
    S  = T_{i_1 ... i_{k-1} i i_{k+1} ... i_n} W_{j_1 ... j_{l-1} i j_{l+1} ... j_n},
\end{equation}
the summation convention was used above---the sum over $i$ ranging from $0$ to $d-1$ was performed. Notice, that  $S$ has $n+m-2$ indices: $i_1,...,i_{k-1},i_{k+1},...,i_n,j_1,...,j_{l-1},j_{l+1},...,j_m$. Tensor $S$  can be graphically represented as
\begin{equation}
    \includegraphics[width=1.\columnwidth]{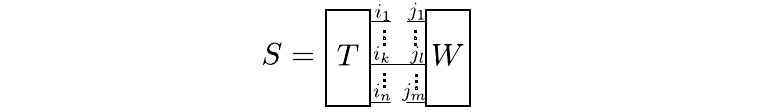}
\end{equation}
Free legs correspond to $n+m-2$ indices of $S$, connected legs correspond to contracted indices .We can combine more tensors into a network, in which each link denotes a contraction of one indices pair. For a more detailed introduction to a tensor networks formalism see Ref. \cite{Orus2014}

Let us now demonstrate the construction of multi-index tensors representing quantum channels, which allow to express a link product using tensor networks formalism. 
Let $C \in \textrm{Lin} \left(\mathcal{H}_1 \otimes \mathcal{H}_2 \otimes ... \otimes \mathcal{H}_N \right)$ be a CJ matrix of a channel whose input and output subspaces are $\mathcal{H}_1, \mathcal{H}_2, ..., \mathcal{H}_N$ (for our further considerations it does not matter which spaces are outputs and which are inputs). Operator $C$ can be written as 
\begin{equation}
    C = C^{i_1 i_2 ... i_N}_{i'_1 i'_2 ... i'_N} \ket{i_1 i_2 ... i_N} \bra{i'_1 i'_2 ... i'_N},
\end{equation}
where indices $i_j, i'_j$ run from $0$ to $d_j-1$,  $d_j = \textrm{dim}  \left( \mathcal{H}_j \right)$; vectors $\ket{0}, \ket{1},...,\ket{d_j-1}$ form o.-n. basis of $\mathcal{H}_j$. We can construct an equivalent representation of $C$ by concatenating indices $i_j$ and $i'_j$ into one index $k_j$, ranging from $0$ to $d_j^2-1$. Then, we obtain an $N$-index tensor $\tilde C$, whose elements are
\begin{equation}
    \tilde C^{k_1 k_2 ...k_N} = C^{i_1 i_2 ... i_N}_{i'_1 i'_2 ... i'_N} \quad \textrm{for} \quad k_j = d_j i_j + i'_j.
\end{equation}
Notice, that for $N=1$, the described procedure corresponds to a matrix vectorization.

Let $\t{E} : \mathcal{L}(\mathcal{H}_1) \rightarrow \mathcal{L} (\mathcal{H}_2 \otimes \mathcal{H}_3)$, $\t{F}: \mathcal{L}(\mathcal{H}_3 \otimes \mathcal{H}_4) \rightarrow \mathcal{L} (\mathcal{H}_5) $ be quantum channels, and $E \in \mathcal{L}(\mathcal{H}_1 \otimes \mathcal{H}_2 \otimes \mathcal{H}_3)$, $F \in \mathcal{L} (\mathcal{H}_3 \otimes \mathcal{H}_4 \otimes \mathcal{H}_5)$ the corresponding CJ operators. We can construct another channel, $\t{G}: \mathcal{L} (\mathcal{H}_1 \otimes \mathcal{H}_4) \rightarrow \mathcal{L} (\mathcal{H}_2 \otimes \mathcal{H}_5)$ by using part of $\t{E}$  ($\mathcal{H}_3$) a part of input of $\t{F}$. Then, by construction, $\t{G} \in \t{Comb} [(\mathcal{H}_1, \mathcal{H}_2), (\mathcal{H}_4, \mathcal{H}_5)] $ , and the corresponding CJ matrix $G$ can be written using link product:
\begin{equation}
    G = E \star F = \t{Tr}_{\mathcal{H}_3} [(E \otimes \openone_{\mathcal{H}_4 \otimes \mathcal{H}_5})(\openone_{\mathcal{H}_1 \otimes \mathcal{H}_2} \otimes F^{T_{\mathcal{H}_3}})]
\end{equation}
Equivalently, we can write down the relation between $\tilde G, \tilde E, \tilde F$ using tensor network formalism:
\begin{equation}
    \includegraphics[width=\columnwidth]{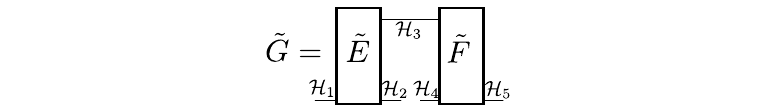}.
\end{equation}

This procedure can be directly generalized to channels with more input and output subspaces. We can also concatenate more channels, to represent quantum combs consisting of many teeth. 
 In the main text, to simplify the notation, we remove $\sim$ symbol, and write the symbols of CJ matrices (e.g. $G$, $E$, $F$) in the tensor network notation. Formally, one should understand this notation as indices contraction in the corresponding tensors $\tilde G, \tilde E, \tilde F$. 

\section{Tensor network based optimization: technical details}
\label{app:ten_net_opt}
Let us present an efficient way to compute $S_k$ matrices which are used to optimize over strategy teeth $P_k$, see \eqref{eq:opt_Pk}. Firstly, to be more precise, let us define $P_k^{[i]}$ as a value of $k$-th tooth of an estimation strategy obtained after $i$ full iterations of an optimization algorithm. To update this value to $P_k^{[i+1]}$, we need to construct a matrix $S_k^{[i+1]}$---this is a notation for a matrix $S_k$ with updated values  of $P_l$ for $l<k$ ($P_{l<k} = P_l^{[i+1]}$) and the values of $P_{l>k}$ from the previous iteration ($P_{l>k} = P_l^{[i]}$). This is related with the order of optimization---we optimize from left to right.   Let us also define $S_{k,+} = \sum_{i=1}^N S_{k,i}$, this means that $S_k = S_{k,0} + S_{k,+}$, see \eqref{eq:Sk_sum}. The matrix $S_{k,0}^{[i+1]}$ can be represented as 
\begin{equation}
\label{eq:S1}
    \includegraphics[width=1.\columnwidth]{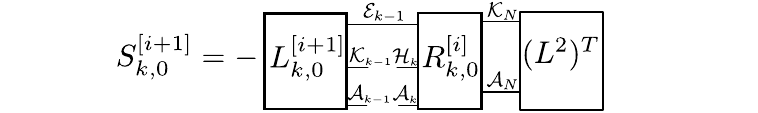},
\end{equation}
where $L_{k,0}^{[i+1]}$, $R_{k,0}^{[i]}$ correspond to contraction of elements from left and right of $P_k$ respectively, and are defined by the following iterative relations:
\begin{equation}
\label{eq:L1}
    \includegraphics[width=1.\columnwidth]{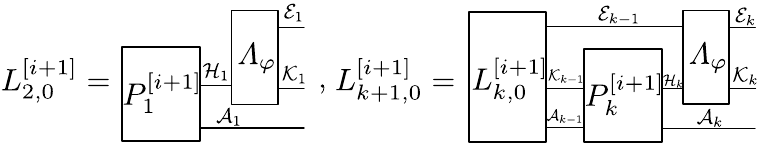},
\end{equation}

\begin{equation}
    \label{eq:R1}
    \includegraphics[width=1.\columnwidth]{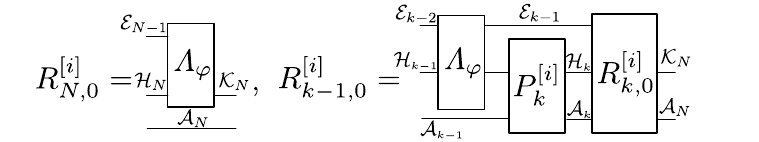}.
\end{equation}
Notice, that we do not define $L_{1,0}$ since there are not any object at the left side of $P_1$.
Instead of calculating matrices $S_{k,i}$ for $i>0$ independently, we can calculate their sum $S_{k,+}$ directly, using the following relation
\begin{equation}
\label{eq:S2}
    \includegraphics[width=1.\columnwidth]{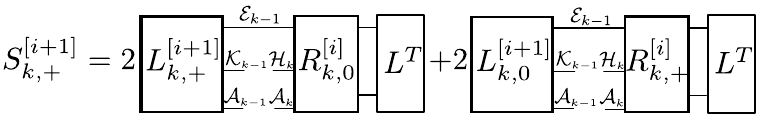},
\end{equation}
where  $L_{k,+}^{[i+1]}$, $R_{k,+}^{[i]}$ are defined by the following iterations:
\begin{equation}
\label{eq:L2}
    \includegraphics[width=1.\columnwidth]{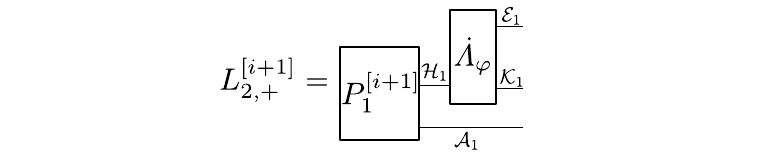},
\end{equation}

\begin{equation}
\label{eq:L3}
    \includegraphics[width=1.\columnwidth]{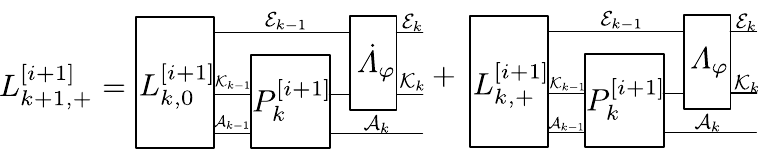},
\end{equation}

\begin{equation}
\label{eq:R2}
    \includegraphics[width=1.\columnwidth]{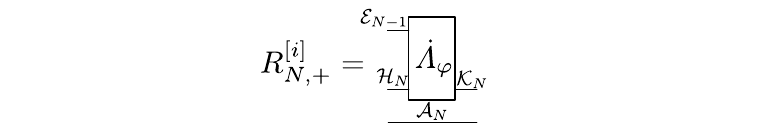},
\end{equation}

\begin{equation}
\label{eq:R3}
    \includegraphics[width=1.\columnwidth]{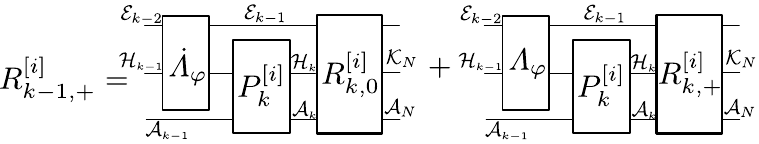}.
\end{equation}
It is straightforward to see that after applying these iterative relations, the figure of merit defined in \eqref{eq:tensor_F} can be indeed written down as \eqref{eq:tensor_PiSi}. Moreover, all tensor networks contractions required to find $P_1^{[i+1]}, P_2^{[i+1]}, ..., P_N^{[i+1]}, L^{[i+1]}$ given $P_1^{[i]}, P_2^{[i]}, ..., P_N^{[i]}, L^{[i]}$   can be done in time $O(N)$. To achieve this, we follow the following algorithm.
\begin{enumerate}
\item Tensors $R_{k,0}^{[i]}$, $R_{k,+}^{[i]}$ are computed using \eqref{eq:R1}, \eqref{eq:R2},  \eqref{eq:R3} for $k \in \{1,2,...,N\}$. When the iterative relations are applied directly, then the time of computation of all of them is $O(N)$.
\item The optimization over $P_1$ is performed, and $P_1^{[i]}$ is replaced with $P_1^{[i+1]}$. Then, $L_{2,0}^{[i+1]}$ and $L_{2,+}^{[i+1]}$ are computed using \eqref{eq:L1} and \eqref{eq:L2}. This allows us to compute $S_{2,0}^{[i+1]}$ and $S_{2,+}^{[i+1]}$ using \eqref{eq:S1} and \eqref{eq:S2}. Then, we perform optimization over $P_2$ and find $P_2^{[i+1]}$, which is used to calculate $L_{3,0}^{[i+1]}$, $L_{3,+}^{[i+1]}$ with the help of \eqref{eq:L1} and \eqref{eq:L3}. This procedure is continued until we find $P_N^{[i+1]}$. Notice, that the time of computation of $L_{k+1,0}^{[i+1]}$, $L_{k+1,+}^{[i+1]}$ using  $L_{k,0}^{[i+1]}$, $L_{k,+}^{[i+1]}$ does not depend on $N$. Therefore, the time of all contractions and optimizations described in this step is $O(N)$.
\item Finally, we find $L^{[i+1]}$ using a standard procedure, described, for example, in \cite{Macieszczak2013, Chabuda2020}.
\end{enumerate}

\section{Optimal metrological protocols for select examples}

\subsection{Perpendicular dephasing}
\label{app:perp_deph_details}

Here we describe the almost optimal protocol for perpendicular dephasing, $N=3$ and $d_\mathcal{A}=2$. Let us start with the optimal protocol for $N=2$. We take as an input state $\ket{\phi^+} = \frac{1}{\sqrt{2}}(\ket{00}_{\mathcal{H}\mathcal{A}}+ \ket{11}_{\mathcal{H}\mathcal{A}})$. After the action of the channel this becomes a mixture of
\begin{equation}
   \frac{1}{\sqrt{2}}(\ket{00} + e^{i\varphi} \ket{11} ), \quad
   \frac{1}{\sqrt{2}}(\ket{10} + e^{i\varphi}\ket{01} ),
\end{equation}
with probabilities $p$ and $1-p$ respectively. Then the error can be detected by projecting the state on one of the two subspaces $\mathcal{C}=\mathrm{span}\{\ket{00}, \ket{11}\}$ and $\mathcal{E}=\mathrm{span}\{\ket{01}, \ket{10}\}$ and then corrected. The action of the second channel leads to the mixture of
\begin{equation}
   \frac{1}{\sqrt{2}}(\ket{00} + e^{i2\varphi} \ket{11} ), \quad
   \frac{1}{\sqrt{2}}(\ket{10} + \ket{01} ),
\end{equation}
with probabilities $p$ and $1-p$ respectively. This state gives QFI at level $4p$ which is an optimal value for $p \ge 0.5$. For $p \le 0.5$ the occurrence of $\sigma_x$ error is more probable than not thus it is more beneficial to treat nonoccurence of $\sigma_x$ as an error and correct the state when it is in span$\{\ket{00}, \ket{11}\}$. Therefore finally we get  $F^{(2)}_Q = 2(1+|1-2p|)$.

Our protocol for $N=3$ does everything as in the case of $N=2$ and then after the action of the second channel it transforms all states in $\mathcal{E}$ (or $\mathcal{C}$ for $p<0.5$) into:
\begin{equation}
    \frac{1}{2}(\ket{0}+\ket{1})\otimes(\ket{0}+e^{-i\pi/4}\ket{1}).
\end{equation}
Numerical computations showed that this results in a QFI that is at worst 8\% percent smaller than the optimal protocol for $d_\mathcal{A}=2$.

Note, that in case when signal comes before a noise the matrix $U_\varphi$ acts on a state immediately after it was corrected. Thus it can be ensured that this state will always be pure which allows for a complete eradication of a noise and $F_Q^{(N)}=N^2$.

\subsection{Parallel dephasing---adaptive vs. entangled based strategy analysis}\label{app:par_deph_adapvspara}
\begin{figure}[ht]
   \includegraphics[width=0.95\columnwidth]{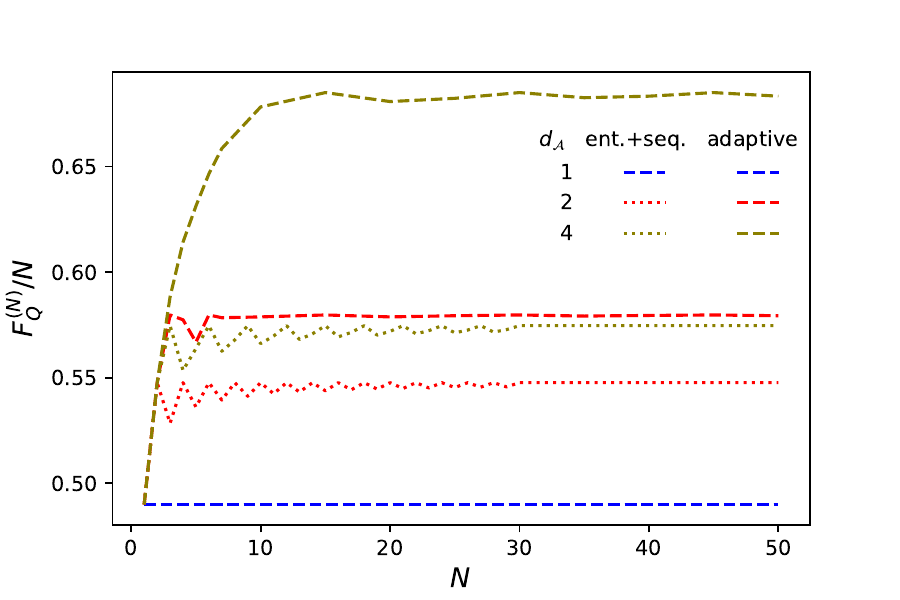}
   \caption{Optimal values of QFI per channel ($F_Q^{(N)}/N$) as a function of the number of channel uses ($N$). General adaptive strategies (dashed line) and strategies involving entanglement and simple sequentiality (dotted line) are compared for different total number of entangled qubits: $1$ (blue, $d_\mathcal{A} = 1$, for this case dashed and dotted line overlap),  $2$ (red, $d_\mathcal{A} = 2$) and $3$ (yellow, $d_\mathcal{A} = 4$). For both cases we allow for splitting $N$ channel uses into $k$ experiments with $n_i$ channel uses ($\sum^k_{i=1} n_i = N$). 
    }
   \label{fig:par_vs_add_app}
\end{figure}

In this Appendix, we present the data showing the comparison between entanglement based and adaptive strategies utilizing the same total number of entangled qubits, see Fig. \ref{fig:par_vs_add_app}. 
For general adaptive strategies (dashed lines), we probe the channels sequentially, and allow for arbitrary control operations in between.
For entanglement based strategies, we do not allow for any intermediate control, but we can optimally choose the input entangled state. Moreover, in the latter case, channels can be probed sequentially, which means that each qubit of an input can be evolved through a number of probed channels. In both cases, we allow for the same total number of channel uses and for the same number of entangled qubits. Notice that $d_\mathcal{A} = 2^m$ ($m$ ancillary qubits) in adaptive strategy correspond to $m+1$ entangled qubits in total.  In some cases, it is advantageous to split $N$ probed channels into subsets of $n_1, n_2, ..., n_k$ channels, and repeat the protocol $k$ times, in $i$-th use, $n_i$ channels are utilized. The optimal division of channels into different coherent runs of the protocol is allowed in both considered cases. 

\subsection{Perpendicular amplitude damping}
\label{app:perp_damp_details}
Let us describe in more details the optimal estimation protocol introduced in Section \ref{sec:perp_damp}, and calculate the associated FI . For a moment, let us assume that after $i$th action of a channel $\Lambda_\varphi$ the probe qubit is in a state
\begin{equation}
\label{eq:rho_phi_app}
\rho^{(i)}_\varphi = \ket{\psi}_{c_i \varphi} \bra{\psi} + \mathcal{O}(\varphi^3),
\end{equation}
where
\begin{equation}
    \ket{\psi}_{c_i \varphi} = e^{-\frac{i}{2} c_i \varphi \sigma_z} \ket{-} =  \cos \left(\frac{c_i \varphi}{2} \right) \ket{-} - i \sin \left(\frac{c_i \varphi}{2} \right) \ket{+},
\end{equation}
the ancillary qubit is in a state $\ket{0}$. In the calculation, we neglect higher order in $\varphi$, since the estimation around $\varphi = 0$ is considered. Using identity
\begin{align}
\label{eq:Vi_action_app}
    &V_i \ket{\psi}_{c_i \varphi} \otimes \ket{0} = \cos \left(\frac{c_i \varphi}{2} \right) \ket{-} \otimes \ket{0} + \\ &- i t_i \sin \left(\frac{ c_i \varphi}{2} \right) \ket{+} \otimes \ket{0} -i \sqrt{1-t_i^2} \sin \left(\frac{c_i \varphi}{2}\right) \ket{-} \otimes \ket{1} 
\end{align}
and \eqref{eq:rho_phi_app}, we can write the output of the control operation as
\begin{equation}
    V_i (\rho_\varphi^{(i)} \otimes \ket{0}\bra{0}) V_i^\dagger = \ket{\chi_i} \bra{\chi_i} +\mathcal{O}(\varphi^3),
\end{equation}
where
\begin{equation}
    \ket{\chi_i} = \ket{\psi}_{c_i t_i \varphi} \otimes \ket{0} - i \sqrt{1-t_i^2} \frac{c_i \varphi}{2} \ket{-} \otimes \ket{1}.
\end{equation}
The ancillary qubit of this output is measured in a computational basis 
 \begin{itemize}
     \item {With probability $p_{1,i} = \frac{c_i^2 \varphi^2}{4} (1-t_i^2) + \mathcal{O}(\varphi^3)$ we measure ancillary qubit in state $\ket{1}$ . Then the 1st qubit is in a state $\ket{-}$ and carries no information about $\varphi$. We can start the whole protocol again.}
     \item{With probability $p_0 = 1- \mathcal{O}(\varphi^2)$ we measure ancillary qubit in $\ket{0}$, and the 1st qubit is then in a state $\ket{\psi}_{c_i t_i \varphi} \bra{\psi} + \mathcal{O}(\varphi^3)$}
 \end{itemize}
 In the 2nd case, we use the output state as input to next channel $\Lambda_\varphi$---the output of $\Lambda_\varphi$ can be then calculated using \eqref{eq:per_damp}, and after expanding around $\varphi=0$, we obtain
 \begin{equation}
     \Lambda_\varphi( \ket{\psi}_{c_i t_i \varphi} \bra{\psi}) = \ket{\psi}_{c_{i+1}  \varphi} \bra{\psi} + \mathcal{O}(\varphi^3),
 \end{equation}
 
  where \begin{equation} \label{eq:ci_rec}c_{i+1} = c_i t_i \sqrt{p} + 1.\end{equation}
  This justifies our initial assumption about the form of the input state of $V_i$---we start with a state $\ket{-} = \ket{\psi_0}$, and during the whole protocol the state is in the form $\ket{\psi}_{c_i \varphi} \bra{\psi} + \mathcal{O}(\varphi^3)$.  The coefficients $c_i$ are given by the recursive relation \eqref{eq:ci_rec} with an initial condition $c_0 = 0$.

  The total probability of a protocol termination in step $i$ is 
  \begin{equation}
      p_i = (1-p_{1,1})...(1-p_{1,i-1}) p_{1,i} = \frac{c_i^2 \varphi^2}{4} (1-t_i^2) + \mathcal{O}(\varphi^4),
  \end{equation}
note that it is equal to $p_{1,i}$ when terms of order $\varphi^4$ and higher are neglected. With probability $1 - \mathcal{O}(\varphi^2)$ the protocol will not terminate after $N$ channel uses, and at the output we will obtain a probe state $\ket{\psi}_{c_N \varphi}\bra{\psi} + \mathcal{O}(\varphi^3)$  and ancillary qubit in state $\ket{0}$. The QFI associated with this output state is
\begin{equation}
    F_\t{out} = c_N^2,
\end{equation}
the measurement which allows to achieve the classical FI equal to QFI is, for example, the measurement in $\ket{\pm}$ basis.

The total FI achieved in the described protocol is
\begin{equation}
\label{eq:FN_perp_strategy}
    F^{(N)} = \sum_{i=1}^{n-1} \frac{\dot p_i^2}{p_i} + F_\t{out} = \sum_{i=1}^{n-1} c_i^2 (1-t_i^2) + c_N^2
\end{equation}
To get an optimal performance, we need to optimize the result over $\{t_i\}$. This can be done step by step---for a given $c_i$ we pick the value $t_i$ that maximizes the sum of the FI associated with the measurement of ancillary system after $V_i$ and the QFI of the probe output state. Therefore, we find $t_i$ maximizing the function
\begin{equation}
    \frac{\dot p_i^2}{p_i} + c_{i+1}^2 = c_i^2(1-t_i^2) + (c_i t_i \sqrt{p}+1)^2
\end{equation}
This is a quadratic function, and it can be easily shown that the optimal choice of $t_i$ is 
\begin{equation}
\label{eq:t_it}
      t_i = \left\{ \begin{matrix} \frac{\sqrt{p}}{c_i (1-p)},~~\t{when} \quad  \frac{\sqrt{p}}{ (1-p)} \le c_i \\ 1 ~~\t{othwerwise}\end{matrix} \right.
  \end{equation}
  After inserting this to \eqref{eq:ci_rec}, we obtain
  \begin{equation}
  \label{eq:c_it}
      c_{i+1} = \left\{ \begin{matrix} \frac{1}{1-p},~~\textrm{when} \quad  \frac{\sqrt{p}}{ (1-p)} \le c_i \\ c_i \sqrt{p} + 1~~ \textrm{othwerwise}\end{matrix} \right.
  \end{equation}
  Notice, we always have $c_i \le c_i \sqrt{p} + 1 \le \frac{1}{1-p}$ for $c_i \le \frac{\sqrt{p}}{ (1-p)}$, which means that $c_i \le c_{i+1}$ and $c_i \le \frac{1}{1-p}$ for any $i$. In the first stage of the protocol (first $k$ channel uses, see Fig. \ref{fig:perp_damping_strategy}) the optimal value of $t_i$ is 1, which means that no error correction is performed. In that stage, $c_i$ keeps growing. Eventually, for $i=k$,  $c_k$ becomes larger than $\frac{\sqrt{p}}{ (1-p)}$, and then $t_k<1$ must be picked, according to \eqref{eq:t_it}. Moreover, according to \eqref{eq:c_it}, it means that $c_{k+1} = \frac{1}{1-p}$. Since now, protocol is in its stable phase---it can be easily seen from \eqref{eq:c_it} and \eqref{eq:t_it}, that $c_i = \frac{1}{1-p}$ and $t_i = \sqrt{p}$ for $i \ge k+1$. The error correction with this value of $t_i$ keeps the probe state in a state $\ket{\psi}_{\varphi/(1-p)}$ as long as ancillary qubit is measured in $\ket{0}$. Then, the FI increase associated with each new channel is $\frac{1}{1-p}$, which is an optimal asymptotic value of QFI per channels for this noise, as a result of bounds derived in Ref. \cite{Kurdzialek2023}. 

\begin{figure}[t]
    \includegraphics[width=0.85\columnwidth]{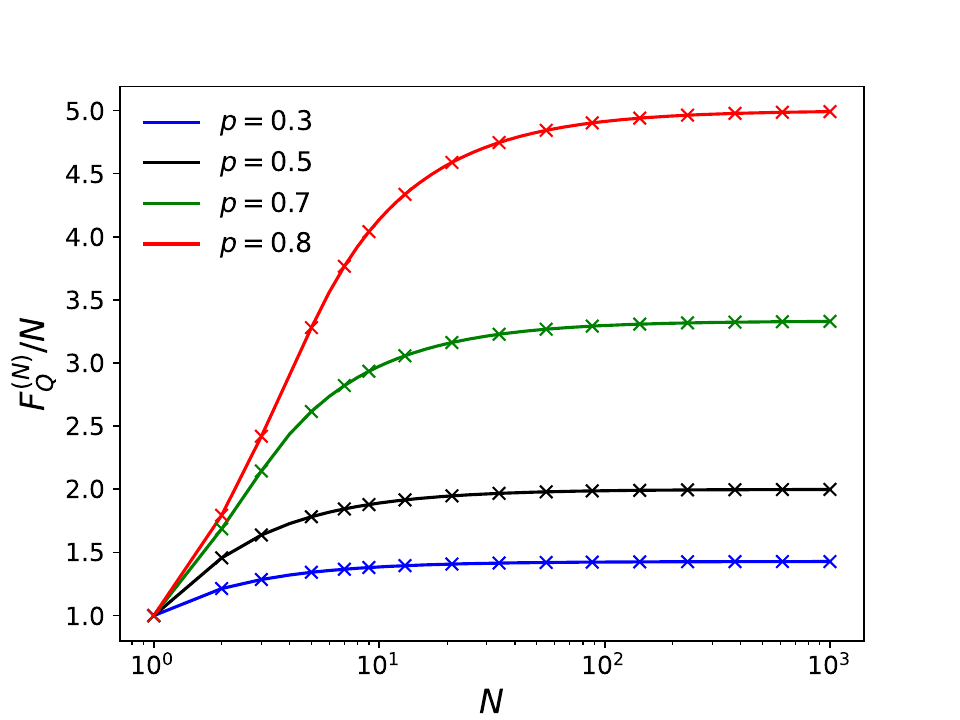}
    \caption{The QFI per channel ($F_Q^{(N)}/N$) as a function of number of channels $N$ for perpendicular amplitude damping noise for different noise parameters $p$. The fundamental bound (solid line) is compared with the FI achieved by the described protocol (crosses). The bound is saturable using this protocol, so crosses coincide with solid lines.  }
    \label{fig:app_perp_damp}
\end{figure}

  To show that this protocol is indeed optimal, we calculated the FI analytically using \eqref{eq:c_it} , \eqref{eq:FN_perp_strategy}, and compared the result with the fundamental bound. The bound turned out to be saturated in all the cases, see Figure \ref{fig:app_perp_damp}.  
  \subsection{Parallel amplitude damping}
\label{app:par_damp_details}
Here we provide details of the optimal strategies for parallel damping, with restriction to 1 or 2 qubits of ancilla, as introduced in Section \ref{sec:parallel_damping}. We provide explicit numerical values for strategies for $p=0.5$, as a comparison with the strategy presented in \cite{Liu2023}.

\subsubsection{1-qubit ancilla}
The Kraus operators $\{L_{i, j}\}$ of the $i$-th tooth of the strategy are as follows. The first tooth prepares an entangled state:
\begin{equation}
    L_{0, 0} = |\psi_0\rangle, \\
\end{equation}
where $| \psi_0\rangle = a_1 |00\rangle + a_2 |11 \rangle $ and for $p=0.5 \, \, \, a_1=0.678, a_2=0.735$. The second tooth's Kraus operators are as follows:
\begin{equation}
\begin{aligned}
    L_{1, 0} &= U (|00\rangle + |11\rangle) (\langle00| + \langle11|),\\
    L_{1, 1} &= |\psi_1\rangle \langle 01|,\\
    L_{1,2} &= |10\rangle \langle 10|,
\end{aligned}
\end{equation}
where $|\psi_1\rangle = b_1|00\rangle + b_2|11\rangle$ and for $p=0.5:$ $b_1=0.590, b_2 = 0.800 + 0.105i$, $U =  0.959|00\rangle\langle00| + (-0.282-0.002i)|00\rangle\langle11| + (0.278+0.047i)|11\rangle\langle00| + (0.945+0.166i)|11\rangle\langle11|$.

\subsubsection{2-qubit ancilla}

\begin{figure}[h!]
    
    \includegraphics[width=0.85\columnwidth]{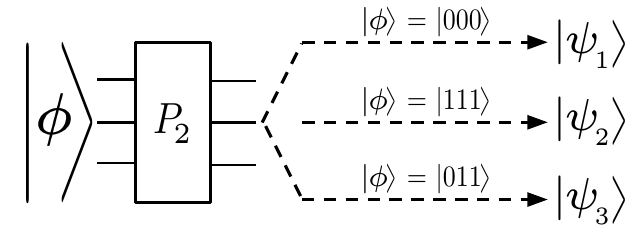}
    \caption{Second tooth, $P_2$, of the optimal strategy for parallel damping noise model with two uses of channel $\Lambda_\varphi$, when ancilla is restricted to 2 qubits. After a state $|\psi_0\rangle$ is prepared by the first tooth and after the action of signal and noise, the 3-qubit input state to $P_2$ is $|\phi\rangle$. $P_2$, acts as follows: if $|\phi\rangle$ is $|000\rangle$, $|\psi_1\rangle$ is prepared; if $|\phi\rangle$ is $|111\rangle$, $|\psi_2\rangle$ is prepared; if $|\phi\rangle$ is $|011\rangle$, $|\psi_3\rangle$ is prepared.}
    \label{fig:par_damping_2qubit_ancilla}
\end{figure}

 For this ancilla size the adaptive bound from \cite{Kurdzialek2023} is saturated. The Kraus operators $\{L_{i, j}\}$ of the $i$-th tooth of the optimal strategy are as follows. The first tooth prepares an entangled state:
\begin{equation}
    L_{0,0} = |\psi_0\rangle, \\
\end{equation}
where $| \psi_0\rangle = a_1 |000\rangle + a_2 |111 \rangle $ and for $p=0.5: \, \, \, a_1=0.675, a_2=0.738$. The second tooth's action is represented schematically in Fig. \ref{fig:par_damping_2qubit_ancilla}. The second tooth's Kraus operators are as follows:
\begin{equation}
\begin{aligned}
    L_{1, 0} &= |\psi_1\rangle \langle000| + |\psi_2\rangle \langle111|,\\
    L_{1, 1} &= |\psi_3\rangle \langle 011|,\\
    L_{1, k} &= |\psi_k\rangle \langle \psi_k| \,\, \mathrm{for} \,\, \psi_k \in \{|001\rangle, |010\rangle, |100\rangle, |101\rangle, |110\rangle \},
\end{aligned}
\end{equation}
where:
\begin{equation}
\begin{aligned}
   |\psi_1\rangle &= b_1|000\rangle + b_2 |101\rangle,\\
   |\psi_2\rangle &= c_1|000\rangle + c_2 |101\rangle,\\
   |\psi_3\rangle &= e_1|000\rangle + e_2|010\rangle + e_3|011\rangle + e_4|101\rangle \\ &+ e_5|110\rangle + e_6|111\rangle
\end{aligned}
\end{equation} 
For $p=0.5:$ $b_1=-0.956,  b_2 = -0.160-0.248i, c_1=0.295, c_2=-0.519-0.803i, e_1=0.062, e_2=-0.041+0.519i, e_3 = 0.289+0.237i, e_4=0.042+0.065i, e_5=0.370+0.247i, e_6=-0.580+0.225i$.

\end{document}